\documentclass[journal]{IEEEtran}
\IEEEoverridecommandlockouts

\usepackage{comment}
\usepackage{caption}

\usepackage[noend]{algpseudocode}
\usepackage{varwidth}
\usepackage{amsmath}

\usepackage{hyperref}
\hypersetup{colorlinks=true,citecolor = blue}

\usepackage{enumerate} 

\newtheorem{definition}{Definition} 

\newtheorem{property}{Property}
\newtheorem{strategy}{Strategy}    
\usepackage{graphicx,epstopdf,algpseudocode,caption,url}   
\usepackage{multirow}  
\graphicspath{{./images}}

\usepackage{stfloats} 

\usepackage{amssymb} 
\usepackage{array}    
\usepackage{arydshln} 
\usepackage{graphics}
\usepackage{color}   
\usepackage{graphicx}   
\usepackage{graphicx,epstopdf,caption,url}   
\usepackage{multirow}  
\newenvironment{proof}{\begin{IEEEproof}}{\end{IEEEproof}}  
\usepackage[ruled,linesnumbered]{algorithm2e}

\ifCLASSINFOpdf

\else

\fi

\begin{document}
%
\title{Temporal Fuzzy Utility Maximization with Remaining Measure}

\author{
	
	Shicheng Wan, Zhenqiang Ye,  Wensheng Gan,~\IEEEmembership{Member,~IEEE} and Jiahui Chen\\   
	
	\thanks{This research was supported in part by the National Natural Science Foundation of China (Grant Nos. 61902079 and 62002136), Guangzhou Basic and Applied Basic Research Foundation (Grant Nos. 202102020928 and 202102020277), National Key-Research and Development Program of China (Grant No. 2020YFB2104003), and the Young Scholar Program of Pazhou Lab (Grant No. PZL2021KF0023).}
	
	\thanks{Shicheng Wan, Zhenqiang Ye, and Jiahui Chen are with the Department of Computer Science, Guangdong University of Technology, Guangzhou 510006, China. (E-mail: scwan1998@gmail.com, yzq66f@gmail.com, csjhchen@gmail.com)}
	
	\thanks{Wensheng Gan is with the College of Cyber Security, Jinan University, Guangzhou 510632, China; and also with Pazhou Lab, Guangzhou 510330, China. (E-mail: wsgan001@gmain.com)}  
	
	\thanks{Corresponding author: Wensheng Gan and Jiahui Chen}
}

\maketitle


\begin{abstract}

High utility itemset mining approaches discover hidden patterns from large amounts of temporal data. However, an inescapable problem of high utility itemset mining is that its discovered results hide the quantities of patterns, which causes poor interpretability. The results only reflect the shopping trends of customers, which cannot help decision makers quantify collected information. In linguistic terms, computers use mathematical or programming languages that are precisely formalized, but the language used by humans is always ambiguous. In this paper, we propose a novel one-phase temporal fuzzy utility itemset mining approach called TFUM. It revises temporal fuzzy-lists to maintain less but major information about potential high temporal fuzzy utility itemsets in memory, and then discovers a complete set of real interesting patterns in a short time. In particular, the remaining measure is the first adopted in the temporal fuzzy utility itemset mining domain in this paper. The remaining maximal temporal fuzzy utility is a tighter and stronger upper bound than that of previous studies adopted. Hence, it plays an important role in pruning the search space in TFUM. Finally, we also evaluate the efficiency and effectiveness of TFUM on various datasets. Extensive experimental results indicate that TFUM outperforms the state-of-the-art algorithms in terms of runtime cost, memory usage, and scalability. In addition, experiments prove that the remaining measure can significantly prune unnecessary candidates during mining.

\textit{Impact Statement}---This article proposes a method to find profitable fuzzy itemsets from temporal databases. The novel algorithm can play an important role in marketing and business, where these discovered patterns are paramount for strategic decision-making. The TFUM algorithm achieves this by integrating fuzzy set theory into itemset mining and then making it possible to transform linguistic terms into quantitative values and making binary machines find both interesting and explicable patterns, such as a young and tall man, beautiful flowers, and sauna days. The proposed remaining measure addresses the ``combinatorial explosion'' problem in a certain way and achieves state-of-the-art performance on various databases (including sparse and dense). TFUM can provide a contribution to the explainable artificial intelligence system, pattern recognition, communication of information applications, and so on.

\end{abstract}


\IEEEpeerreviewmaketitle

\section{Introduction}
\label{sec:introduction}

Data mining technology can be regarded as an algorithmic process that takes data as input and yields useful results as output. It is meaningful to identify and isolate relationships among the different items that may be hidden in massive data, since the first and well-known association rule mining (ARM) algorithm was proposed \cite{agrawal1994fast}. In market basket analysis, the database records items purchased by a customer at a single time as a transaction. The association rule mining task aims to find out the ``association" between sets of items with some strong specified confidence. An example of such an association is the statement that 88\% of transactions involving men from 30 to 39 years old will buy a car. The number 88\% shows the confidence value of the rule. The antecedent of this rule consists of men aged 30 to 39 years old, and the consequent consists of the car alone. This rule means a man has a high probability of purchasing a car when he is between 30 and 39 years old. In the meantime, frequent itemset mining (FIM) \cite{han2000mining,han2004mining,pei2007h} started as a phase in the discovery of association rules, but has been generalized independent of these to many other patterns. In recent years, Aggarwal \textit{et al.} \cite{aggarwal2018spatio} proposed a spatio-temporal FIM algorithm on web data. However, a fundamental limitation of both FIM and ARM is that they both assume that each item cannot appear more than once in each transaction and that all items have the same importance (e.g., weight, unit profit, or risk). In other words, bread will be regarded as interesting, but diamonds will not, because the sale volume of the latter one is far less than that of the prior. The frequency of an itemset cannot be a sufficient indicator of interest in some cases. As a consequence, traditional high-utility itemset mining (HUIM) algorithms discover more realistic and important knowledge than FIM by taking into account non-binary occurrences of items within transactions and various interests in distinct items. The discovered results of traditional HUIM are more interpretable than those of FIM. After decades of development, plenty of algorithms \cite{liu2012mining,gan2020tophui,gan2018survey,wan2022discovering} were proposed and have been applied in many applications like cross-marketing \cite{liu2013mining}, click stream analysis \cite{chu2008efficient}, and traffic management \cite{hai2011moving}.

However, a critical drawback of previous studies is that few of them consider the temporal aspect of databases. If a customer visited the time of a retail last year, his/her transaction record may be invalid at present, and thus be useless for analysis nowadays \cite{wan2022fast}. More generally, consider that transactions in a retail store contain the following five items: \{\textit{apple}\}, \{\textit{ice-cream}\}, \{\textit{neckerchief}\}, \{\textit{sweater}\}, and \{\textit{watermelon}\}. Items such as \{\textit{watermelon}\} and \{\textit{ice-cream}\} are typically best-selling during the summer, whereas items such as \{\textit{sweater}\} and \{\textit{neckerchief}\} sales better in the winter than in the summer, and \{\textit{apple}\} will be needed all year round. Therefore, in this case, traditional HUIM approaches are unable to deal with temporal transaction databases. In recent years, Chen \textit{et al.} \cite{chen2022shelf} proposed a novel on-shelf utility mining algorithm and then discovered maximal profitable product combinations by considering on-shelf periods of items. However, all the above data mining algorithms simply reveal that discovered itemsets (i.e., product combinations) are profitable but not capable of reflecting quantization information. The Likert scale is a notable example of this issue. There are five different types of answers: totally disagree, disagree a little, neutral opinion, agree a little, and totally agree. The fact remains that such imprecisely defined "extents/classes" play an important role in human thinking, particularly in the domains of pattern recognition, communication of information, and abstraction. Hence, how to make a machine learns linguistic representations (e.g., sweet, hot, and beautiful) is an interesting but challenging task. Huang \textit{et al.} \cite{huang2017temporal} figured out that fuzzy quantifiers can be a fuzzy approximation to linguistic representations. By adopting fuzzy set theory \cite{zadeh1965fuzzy} and user-defined membership functions, users can easily comprehend how many products are sold at an expensive, moderate, or cheap price. The characteristic of fuzzy set is that determining whether an element belongs to a set cannot be answered simply by ``yes'' or ``no'', whereas it is gradient. Huang \textit{et al.} \cite{huang2017temporal} first defined the temporal fuzzy utility itemset mining task. That is, given a predefined minimum threshold and user-defined membership functions, mining a complete set of high temporal fuzzy utility itemset from temporal quantitative transaction databases. The aim of the novel mining task is to find patterns that are easily interpretable by users, and thus can help in understanding the results. However, their proposed algorithm is a ``generate-and-test/two-phase'' approach and thus performs poorly. Then, Hong \textit{et al.} continuously proposed several new studies \cite{hong2020using,hong2022one} to improve the performance of algorithms in the temporal fuzzy utility itemset mining domain.

Recently, Ye \textit{et al.} \cite{ye2022fuzzy} proposed an effective temporal fuzzy utility itemset mining algorithm, which performs better than the algorithm of Hong \textit{et al.} \cite{hong2022one}. Though their approach utilizes list structure to efficiently store heuristic information, the adopted upper-bounds are too loose to consume a large amount of runtime and memory. This issue motivates us to further develop the temporal fuzzy utility itemset mining. In this paper, we propose a new one-phase list-based algorithm called TFUM. The major contributions of this paper are summarized as follows.

\begin{itemize}	
	\item A novel structure, called \textit{RTF-List}, is proposed. We first integer maximal remaining fuzzy utility into temporal fuzzy-lists and propose a remaining measure to decide whether the fuzzy itemsets should be pruned or not.
	
	\item  Without checking transaction identifications one by one, we propose a TP-table which is capable of fast locating quantitative transactions by period identification.
	
	\item  The novel algorithm holds the downward closure property and adopts the temporal fuzzy utility upper-bound ratio to prune the search space as far as possible.
	
	\item Extensive experiments have been conducted on four datasets (including sparse and dense). Under various periods and minimal thresholds, the experimental results reveal that the novel proposed algorithm performs far better than the state-of-the-art algorithms in terms of run time, memory consumption, and scalability.	
\end{itemize}

The rest of the content in this paper is organized as follows. The related work will be briefly reviewed in Section \ref{sec:relatedWork}. Section \ref{sec:preliminaries} introduces some basic definitions and the problem statement. Section \ref{sec:algorithm} describes the proposed algorithm, and Section \ref{sec:example} illustrates a detailed example of our approach. Finally, extensive experiments and responding analysis are shown in Section \ref{sec:experimental}, and future work is finally presented in Section \ref{sec:conclusion}.
 
\section{Related Work}  \label{sec:relatedWork}

\subsection{Itemset Mining Using List Structure}

Based on the way of discovering interesting patterns from databases, many list-based algorithms in itemset mining domain actually belong to a one-phase approach. In high utility itemset mining (HUIM), HUI-Miner \cite{liu2012mining} is the earliest and most famous list-based algorithm. Compared to the two-phase HUIM algorithms \cite{yao2004foundational, liu2005two}, HUI-Miner is a pattern-growth approach and does not need to scan the original database more than twice. It firstly finds promising low-level itemsets (which are potential high utility itemsets) and then constructs utility-lists for these itemsets. A utility-list consists of several tuples, and there are two basic elements (i.e., identifiers of the transaction involving the itemset and the utility of the itemset) in each tuple. Then, the algorithm joins different utility-lists to generate new utility-lists of high-level itemsets until there are no more itemsets generated. After that, numerous optimization algorithms \cite{gan2020tophui, fournier2014fhm, liu2015mining, krishnamoorthy2017hminer, duong2018efficient, krishnamoorthy2019mining, wu2022ubp} have been proposed in the last decades.

Besides, to meet the requirements of various domains, various list-based itemset mining extensions have been proposed, which take constraints and more complex data types into account. Wan \textit{et al.} \cite{wan2021fuim} first adopted fuzzy-lists to mine high-fuzzy itemsets from quantitative transaction databases. They proposed that the remaining fuzzy utility of each fuzzy itemset is a tighter upper bound than FUUB \cite{lan2015fuzzy}. And the extensive experiments reveal the novel upper-bound plays an important role in their algorithm. Furthermore, Cui \textit{et al.} \cite{cui2022fri} considered frequency constraint and utilized fuzzy-lists to find valuable and interesting fuzzy rare itemsets. In transaction databases, high utility itemsets do not provide information about the purchase quantities of items. Then, Nouioua \textit{et al.} \cite{nouioua2021fhuqi} designed the utility lists of $q$-itemsets to solve the issue. Chen \textit{et al.} \cite{chen2021mining} noticed most existing HUIM algorithms assume that itemsets always occur regardless of the period. However, this assumption is not realistic in most cases, such as the sold volume of ice cream. Hence, they considered the rich information (e.g., quantity and period) in databases and adopted utility-lists to discover on-shelf high-utility quantitative itemsets. In addition, recently, Chen \textit{et al.} \cite{chen2021discovering} mined high utility-occupancy patterns from uncertain data by probability-utility-occupancy list and probability-frequency-utility table. Considering the existence probability of items will be more realistic than assuming the items must occur in transactions.

\subsection{Itemset Mining Using Fuzzy Theory}

Since Srikant and Agrawal \cite{srikant1996mining} first proposed a fuzzy mining approach to discover quantitative association rules, they have opened a novel fuzzy pattern mining domain. Though their proposed approach utilizes a naive method (i.e., the generate-and-test mechanism), which leads to inefficiency, the discovered results of the new algorithm can be more convenient for helping decision makers to understand and use. Then, Chan and Au \cite{chan1997mining} figured out that the discovered quantitative association rules by discretizing the domains of quantitative attributes into intervals is not concise and meaningful enough. Therefore, they employed linguistic terms to reveal quantitative association rules by setting membership functions in advance. They called the mining rules as fuzzy association rules, because their proposed algorithm adopts the fuzzy set theory. Later, Kuok \textit{et al.} \cite{kuok1998mining} computed the membership value of a high-level fuzzy itemset by applying a minimum operation to get the overlap value of membership regions in its consisting fuzzy items. Since the user-predefined membership function may divide an item into several membership regions, Hong \textit{et al.} \cite{hong1999mining} noticed that using the maximal membership value as the fuzzy value of the item allows the item to be generalized to a certain extent. This idea also affects many subsequent fuzzy pattern mining algorithms. The study \cite{papadimitriou2005fuzzy} employed a heuristic method and tree structure to mine fuzzy association rules. The experimental results showed that their proposed algorithm performs better than previous Apriori-based fuzzy association rule mining algorithms. Then, there are also several variations on the original FP-Growth algorithm \cite{han2000mining} proposed to discover fuzzy frequent itemsets from quantitative databases \cite{lin2010linguistic, lin2013survey, lin2014mining}. However, frequent itemsets often stand in front of the barrier of discovering frequent but low profit patterns.

As similar to traditional high utility itemset mining, Wang \textit{et al.} \cite{wang2009fuzzy} first proposed fuzzy utility mining (FUM) domain to solve the above issue. They had successfully extended HUIM with fuzzy sets to handle quantitative transaction databases. Subsequently, Lan \textit{et al.} \cite{lan2015fuzzy} proposed another different kind of FUM algorithm named TPFU. TPFU assesses the utility of an item based on both its linguistic terms defined by users and the membership values of terms scoped in a user-predefined membership function. Though TPFU belongs to two-phase approach, it has integrated minimum operation and maximal membership value into the fuzzy utility mining well. Recently, Wan \textit{et al.} \cite{wan2021fuim} proposed a list-based fuzzy utility mining approach called FUIM, and the novel algorithm outperforms TPFU in terms of runtime and memory usage since FUIM is a one-phase approach. However, the time attribute of a discovered pattern plays an important role in many real applications, like weblog analysis, business decisions, and so on. Huang \textit{et al.} \cite{huang2017temporal} therefore defined a novel task of discovering high temporal fuzzy utility itemsets from temporal quantitative databases (temporal fuzzy utility itemset mining, TFUIM). Then, Hong \textit{et al.} \cite{hong2020using,hong2022one} have successively published two TFUIM algorithms (i.e., FHTFUP and ATTFUM) which are implemented by tree structure. In particular, ATTFUM is a one-phase algorithm and adopts an array-embedded tree structure. It utilizes an array-list to keep key information about interesting fuzzy itemset in each tree node. Recently, Ye \textit{et al.} \cite{ye2022fuzzy} adopted \textit{TF-Lists} to compress the temporal quantitative databases. The experiments revealed that the novel algorithm performs better than previous algorithms on dense databases.

\section{Preliminaries and Problem Statement}
\label{sec:preliminaries}

This section firstly introduces some basic and commonly used notations and definitions in this paper. Most of them are proposed from previous studies \cite{huang2017temporal,hong2022one,ye2022fuzzy}. A finite set $I$ = \{$x_1$, $x_2$, $\ldots$, $x_n$\} is consisted of $n$ distinct items. Itemset and transaction are both subsets of $I$. The difference between them is that a transaction can be regarded as an $k$-itemset, which means that it contains $k$ different items (1 $\le$ $k$ $\le$ $n$). In addition, the length of $k$-itemset $X$ is denoted as $|X|$ = $k$. Each transaction is assigned a unique transaction identification (simplified as \textit{Tid}). A period is a set of spatio-temporal data that may contain zero, one, or more transactions ($P$ = \{$T_1$, $\ldots$, $T_m$\}). We also call a temporal quantitative transaction if it is contained in a period, and a transaction cannot belong to two or more different periods at the same time. The fuzzy set is defined by the membership function, and the membership value range is [0, 1]. Fuzzy itemsets that are contained in temporal quantitative transactions are called temporal fuzzy itemsets. Then, a temporal quantitative database \textit{TQD} consists of one or more periods.

\begin{figure}[h]
	\centering
	\includegraphics[scale=0.45]{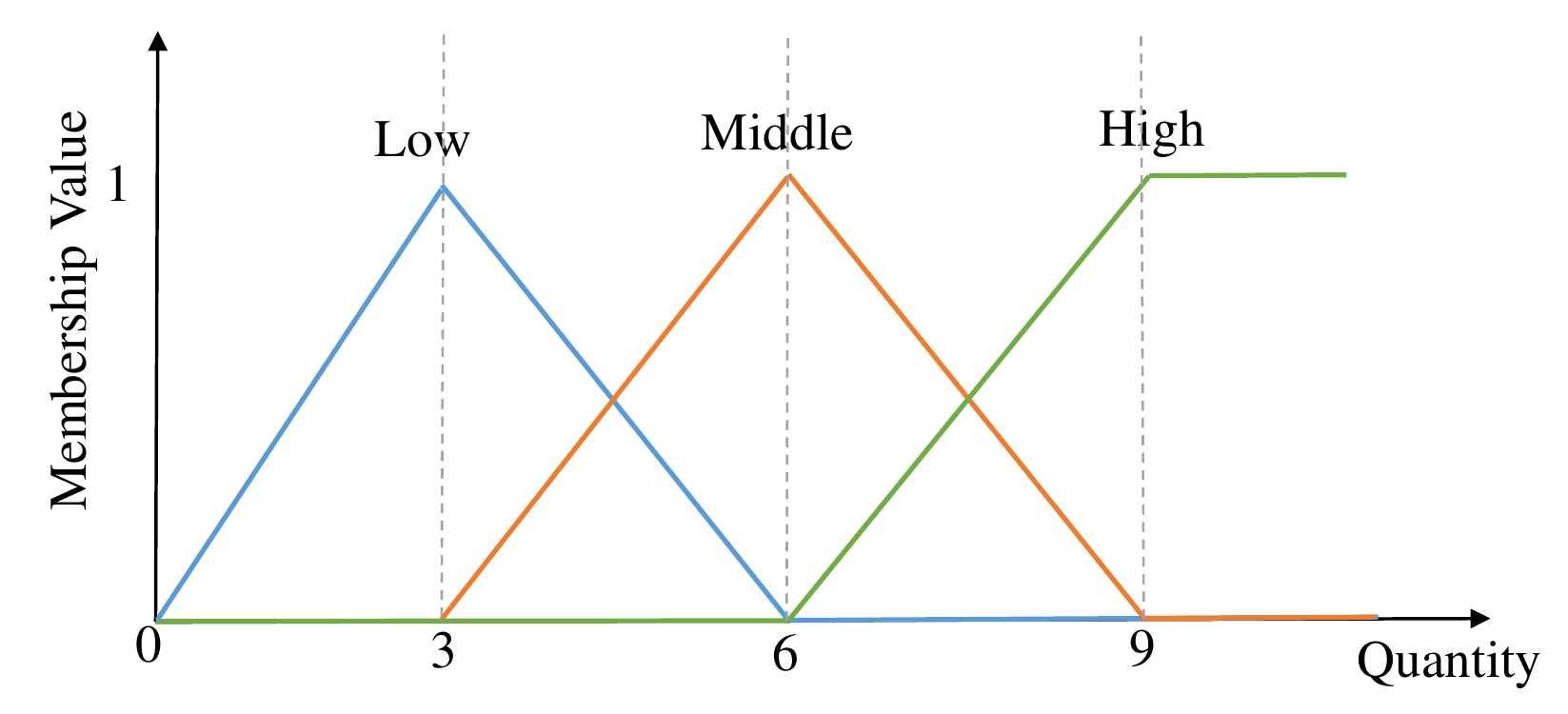}
	\caption{The predefined membership function.}
	\label{fig:defMembership}
\end{figure}

In this paper, we take a temporal quantitative database (Table \ref{tab:database}) and a predefined membership function (Fig. \ref{fig:defMembership}) as our running example. There are six distinct items (i.e., $A$, $B$, $C$, $D$, $E$, and $F$), ten transactions, and five periods in the running example database (\textit{TQD})\footnote{In order to make the expression more concise, we will always use the symbol ``\textit{TQD}'' instead of ``temporal quantitative database'' during discussion.}. The number in each row of \textit{TQD} represents the internal utility (e.g., quantity) of the corresponding item (which is listed in the first row). Furthermore, the external utility (e.g., profit) of each item is shown in Table \ref{tab:profit}. Finally, we will formalize the problem definition for TFUIM.

\begin{table}[!ht]
	\centering
	\caption{A sample temporal quantitative database}
	\label{tab:database}
		\begin{tabular}{cccccccc}
			\hline
			\textbf{Period} & \textbf{\textit{Tid}} &\textbf{\textit{A}} & \textbf{\textit{B}} & \textbf{\textit{C}} & \textbf{\textit{D}} & \textbf{\textit{E}} & \textbf{\textit{F}} \\
			\hline
			$P_1$ & $T_1$ & 1 & 0 & 3 & 0 & 1 & 2 \\
			$P_1$ & $T_2$ & 0 & 4 & 0 & 3 & 0 & 1 \\
			$P_2$ & $T_3$ & 0 & 2 & 0 & 1 & 2 & 0 \\
			$P_2$ & $T_4$ & 2 & 0 & 6 & 1 & 3 & 0 \\
			$P_3$ & $T_5$ & 0 & 1 & 6 & 3 & 0 & 3 \\
			$P_3$ & $T_6$ & 0 & 3 & 0 & 0 & 1 & 7 \\
			$P_4$ & $T_7$ & 2 & 4 & 1 & 5 & 8 & 1 \\
			$P_4$ & $T_8$ & 1 & 6 & 2 & 0 & 1 & 0 \\
			$P_5$ & $T_9$ & 7 & 0 & 0 & 3 & 0 & 0 \\
			$P_5$ & $T_{10}$ & 3 & 1 & 0 & 6 & 1 & 9 \\
			\hline
		\end{tabular}
\end{table}

\begin{table}[h]
	\centering
	\caption{The external utility of each item}
	\label{tab:profit}
	\begin{tabular}{ccccccc}
			\hline
			\textbf{Item} &\textbf{\textit{A}} & \textbf{\textit{B}} & \textbf{\textit{C}} & \textbf{\textit{D}} & \textbf{\textit{E}} & \textbf{\textit{F}} \\
			\hline
			\textbf{Utility} & 9 & 5 & 4 & 2 & 1 & 7 \\
			\hline
	\end{tabular}
\end{table}

\begin{definition}
	\rm As previous content introduced, the number of linguistic terms of an item is depended on the regions of the given membership function. The fuzzy set of an item $x_i$ in a temporal quantitative transaction $T_j$ is defined as $f_{ij}$ = \{$\frac{f_{ij1}}{R_{i1}}$ + $\ldots$ + $\frac{f_{ijl}}{R_{il}}$ + $\ldots$ + $\frac{f_{ijh}}{R_{ih}}$\}. Fuzzy membership degree $f_{ijl}$ is calculated from quantity $q(x_i, T_j)$ and region $R_{il}$.
\end{definition}

For example, consider the \textit{TQD} in Table \ref{tab:database} and the membership function shown in Fig. \ref{fig:defMembership}, since the quantity of $A$ in transaction $T_9$ is 7 that the region of $A$ in $T_9$ are \textit{Middle} and \textit{High}. Based on the membership function, the region values of $R_{A, \textit{Middle}}$ and $R_{A, \textit{High}}$ are 0.67 and 0.33, respectively. Therefore, the fuzzy set of item $A$ in $T_9$ ($f_{A, T_9}$) is \{$\frac{0.67}{A.\textit{Middle}}$ + $\frac{0.33}{A.\textit{High}}$\}.

\begin{definition}
    \label{def:fu}
	\rm The fuzzy utility of $l$-th fuzzy region of an item $x_i$ in a temporal quantitative transaction $T_j$ is defined as \textit{fu}$_{ijl}$($x_i$, $T_j$) = $f_{ijl} \times q(x_i, T_j) \times p(x_i)$. And the fuzzy utility of $x_i$ in \textit{TQD} is the summation of  \textit{fu}$_{ijl}$($x_i$, $T_j$). That is, \textit{fu}$_{il}$($x_i$) = $\sum_{x_i \in T_j \land T_j \in \textit{TQD}}$\textit{fu}$_{ijl}$($x_i$, $T_j$). Besides, the fuzzy utility of a fuzzy itemset $X$ in $T_j$ is denoted as \textit{fu}$_{jX}$($X$, $T_j$) = $f_{jX}$ $\times$ $\sum_{x_i \in X}$[$q(x_i, T_j) \times p(x_i)$], where $f_{jX}$ = \textit{Min}\{$f_{\textit{ijl}} | x_i \in X \land R_{il} \in x_i$\}. Similarly, the fuzzy utility of $X$ in \textit{TQD} is defined as \textit{fu}$_{X}$ = $\sum_{X \subseteq T_j \land T_j \in \textit{TQD}}$\textit{fu}$_{jX}$($X$, $T_j$). In addition, it should be pointed out that different fuzzy regions of an item cannot occur in a same fuzzy itemset in the meanwhile. In other words, an item cannot both \textit{Low} and \textit{Middle} in a fuzzy itemset together.
\end{definition}

For instance, the item $C$ occurs in transactions $T_1, T_4, T_5, T_7$ and $T_8$. The quantity of $C$ in these transactions are 3, 6, 6, 1 and 2, respectively. Based on the Fig. \ref{fig:defMembership} and above definitions, $f_{C, T_8, \textit{Low}}$ is 0.67, and then \textit{fu}$_{C, T_8, \textit{Low}}$($C$, $T_8$) is 0.67 $\times$ 2 $\times$ 4, which is 5.36. By the same way, \textit{fu}$_{C, T_1, \textit{Low}}$($C$, $T_1$) and \textit{fu}$_{C, T_7, \textit{Low}}$($C$, $T_7$) are 12 and 1.32, respectively. In \textit{TQD}, \textit{fu}$_{\textit{C.Low}}$ is 5.36 + 12 + 1.32 = 18.68 and \textit{fu}$_{\textit{C.Middle}}$ is 48. Consider a fuzzy itemset \{\textit{A.Low}, \textit{C.Middle}\} in $T_4$. Since the minimum fuzzy membership degree of \{\textit{A.Low}, \textit{C.Middle}\} is 0.67, and thus \textit{fu}$_{\textit{A.Low}, \textit{C.Middle}}$ is 0.67 $\times$ (2 $\times$ 9 + 6 $\times$ 4) = 28.14.

\begin{definition}
	\rm The fuzzy utility of a $T_j$ in \textit{TQD} is defined as \textit{tfu}$_{j}$ = $\sum_{x_i \in T_j \land f_{ijl} \in f_{ij}}$\textit{fu}$_{ijl}$($x_i$, $T_j$). In addition, the start transaction period \textit{STP}$_{iz}$ of $x_i$ is firstly occurring time period $P_z$ in \textit{TQD} of the transaction ($x_i \in T_j$). The last transaction periods (\textit{LTP}) of an itemset $X$ is a set of periods from the latest start transaction periods of all items in $X$ to the end period of \textit{TQD}. Then, the temporal fuzzy utility ratio of a fuzzy itemset $X$ is formulated as \textit{tfur}$_{X}$ = $\frac{\sum_{X \subseteq T_j \land T_j \in \textit{LTP}_X}\textit{fu}_{jX}}{\sum_{T_j \in \textit{LTP}_X}\textit{tfu}_j}$.
\end{definition}

For example, \textit{STP}$_{\textit{B.Low}}$ is $P_1$ because \textit{B.Low} first appears in $T_2$, and $T_2$ belongs to $P_1$. Take the fuzzy itemset \{\textit{B.Low, C.Middle}\} as an example. \textit{STP}$_{\textit{B.Low}}$ and \textit{STP}$_{\textit{C.Middle}}$ are $P_1$ and $P_6$, respectively. Therefore, the \textit{LTP} of \{\textit{B.Low, C.Middle}\} is a set of \{$P_2$, $P_3$, $P_4$, $P_5$\}.

\begin{definition}
	\rm Given a user-defined minimal utility threshold $\gamma$, a fuzzy itemset $X$ is assumed to be a high temporal fuzzy utility itemset (abbreviated as \textit{HTFUI}) if and only if its \textit{tfur}$_{X}$ is higher than or equal to $\gamma$. Otherwise, $X$ is a low temporal fuzzy utility itemset (\textit{LTFUI}), which is useless for users.
\end{definition}

For example, if $\gamma$ = 20\%\footnote{Unless specifically specified, without loss of generality, we are going to default that $\gamma$ is 20\% in this paper.}, the temporal fuzzy utility ratio of \{\textit{A.Low, F.High}\} is 52.03\% $>$ 20\%, and thus it is an \textit{HTFUI}. Table \ref{tab:HTFUIs} lists other \textit{HTFUIs} when $\gamma$ is 20\%.

\begin{table}[h]
	\centering
	\caption{A set of high temporal fuzzy utility itemsets}
	\label{tab:HTFUIs}
		\begin{tabular}{ll}
			\hline
			\textbf{\textit{HTFUIs}} & \textbf{\textit{tfur}} \\
			\hline
			$\{$\textit{A.Low}, \textit{F.High}$\}$ & 52.03\% \\
			$\{$\textit{A.Middle}$\}$ & 24.4\% \\
			$\{$\textit{D.Low}, \textit{A.Middle}$\}$ & 26.73\% \\
			$\{$\textit{D.Middle}, \textit{A.Low}$\}$ & 22.53\% \\
			$\{$\textit{D.Middle}, \textit{A.Low}, \textit{F.High}$\}$ & 58.97\% \\
			$\{$\textit{D.Middle}, \textit{F.High}$\}$ & 43.36\% \\
			$\{$\textit{F.High}$\}$ & 36.42\% \\
			\hline
		\end{tabular}
\end{table}

\textbf{Problem statement}. Based on previous introduced definitions, given a user-specified minimal utility threshold and a user-defined membership function, we define the problem of temporal fuzzy utility itemset mining (TFUIM) task as discovering the set of high temporal fuzzy utility itemsets in a temporal quantitative database.

\section{The Proposed Flexible Algorithm}
\label{sec:algorithm}

It can be found that \{\textit{A.Middle}\} and \{\textit{D.Low, A.Middle}\} are \textit{HTFUIs}, but \{\textit{D.Low}\} is not in Table \ref{tab:HTFUIs}. Note that the temporal fuzzy utility ratio is neither anti-monotonic nor monotonic. We will discuss how to make our approach hold downward closure property in this section. We also use the temporal remaining fuzzy utility upper-bound to improve TFUM's performance. The details will be introduced in the following.

\subsection{Downward Closure Property}

\begin{definition}
	\rm The maximal fuzzy utility of a fuzzy item $x_i$ in a temporal quantitative transaction $T_j$ is denoted as \textit{mfu}$_{ij}$ = \textit{Max}\{\textit{fu}$_{ij1}$, \textit{fu}$_{ij2}$, $\ldots$, \textit{fu}$_{ijl}$\}. Then, the summation of the maximal fuzzy utility of all fuzzy items in $T_j$ can be a pre-evaluated value of $x_i$. That is, \textit{mtfu}$_{j}$ = $\sum_{x_i \in T_j}$\textit{mfu}$_{ij}$. Obviously, the maximal fuzzy utility of any $x_i$ in $T_j$ is never higher than the maximal fuzzy utility of $T_j$ (i.e., \textit{mfu}$_{ij}$ $\le$ \textit{mtfu}$_{j}$).\footnote{Specially, \textit{mtfu} is the upper-bound of all fuzzy items in corresponding transactions, but \textit{tfu} is not in this paper (as shown in Table \ref{tab:comparison}). Please see the study \cite{hong2022one} for more information on the proof.}
\end{definition}

For example, consider $T_7$ in Table \ref{tab:database}. The quantity of the item $D$ is 5, and then the values for $f_{D, T_7, \textit{Low}}$ and $f_{D, T_7, \textit{Middle}}$ are 0.33 and 0.67, respectively. Hence, \textit{mfu}$_{D, T_7}$ is \textit{Max}\{\textit{fu}$_{D, T_7, \textit{Low}}$, \textit{fu}$_{D, T_7, \textit{Middle}}$\} = \textit{Max}\{0.33 $\times$ 5 $\times$ 2, 0.67 $\times$ 5 $\times$ 2\} = 6.7. The \textit{mfu} of other items in $T_7$ can be calculated with the same method. Therefore, \textit{mtfu}$_{T_7}$ = \textit{fu}$_{A, T_7, \textit{Low}}$ + \textit{fu}$_{B, T_7, \textit{Low}}$ + \textit{fu}$_{C, T_7, \textit{Low}}$ + \textit{fu}$_{D, T_7, \textit{Middle}}$ + \textit{fu}$_{E, T_7, \textit{High}}$ + \textit{fu}$_{F, T_7, \textit{Low}}$ = 12.06 + 13.4 + 1.32 + 6.7 + 5.36 + 2.31 = 41.15.

\begin{definition}
	\rm We defined the start transaction period (\textit{STP}) of all fuzzy items as \textit{STP}$_{\textit{all}}$ = \textit{max}$_{\textit{TP}}$\{\textit{STP}$_1$, \textit{STP}$_2$, $\ldots$, \textit{STP}$_n$\}, where $n$ is the number of fuzzy items, and \textit{max}$_{\textit{TP}}$ operation has the latest time period (\textit{LTP}) of the attached parameters. In addition, \textit{LTP}$_{\textit{all}}$ means all the time periods from \textit{STP}$_{\textit{all}}$ to the last time period of \textit{TQD}.
\end{definition}

For example, the \textit{STP} of \textit{F.High} is $T_{10}$ and $T_{10}$ is the last transaction in \textit{TQD}, that is \textit{STP}$_{\textit{all}} = T_{10}$. Then \textit{LTP}$_{\textit{all}}$ contains only one period $P_5$. 

\begin{definition}
	\rm The temporal fuzzy utility upper-bound ratio of an itemset $X$ is defined as \textit{tfuubr}$_X$ = $\frac{\sum_{X \subseteq T_j \land T_j \in \textit{LTP}_X}\textit{mtfu}_{jX}}{\sum_{T_j \in \textit{LTP}_{\textit{all}}}\textit{tfu}_j}$\footnote{In this paper, we use itemset instead of fuzzy itemset will get a more loose upper-bound than before, but experiments show this case does not affect obtaining the right final results.}. Furthermore, if \textit{tfuubr}$_X$ is higher than or equal to the user-defined minimal threshold $\gamma$, it will be assumed as a high temporal fuzzy utility upper-bound itemset (\textit{HTFUUI}) and should be further checked for its temporal fuzzy utility ratio. Otherwise, it is a \textit{LTFUI}. Hence, the set of \textit{HFUIs} is actually a subset of \textit{HTFUUIs}. The proof details can be referred to the study \cite{huang2017temporal}.
\end{definition}

\begin{table}[h]
	\centering
	\caption{A comparison table (\textit{tfu} vs. \textit{mtfu})}
	\label{tab:comparison}
		\begin{tabular}{cccccc}
		\hline
			\hline
			\textbf{\textit{Tid}} & \textbf{$T_1$} & \textbf{$T_2$} & \textbf{$T_3$} & \textbf{$T_4$} & \textbf{$T_5$} \\
			\textbf{\textit{tfu}} & 29.3 & 28.31 & 8.7 & 39.72 & 52.65\\ 
			\textbf{\textit{mtfu}} & 24.68 & 21.71 & 8.7 & 39.72 & 52.65 \\
			\hline
			\textbf{\textit{Tid}} & \textbf{$T_6$} & \textbf{$T_7$} & \textbf{$T_8$} & \textbf{$T_9$} & \textbf{$T_{10}$} \\
			\textbf{\textit{tfu}} & 64.33 & 53.69 & 38.66 & 69 & 103.98\\ 
			\textbf{\textit{mtfu}} & 48.16 & 41.15 & 38.66 & 48.21 & 103.98 \\
			\hline
			\hline
		\end{tabular}
\end{table}

\begin{property}
	\label{pro:fuzzyDC}
	\rm Considering the definition of the temporal fuzzy utility upper-bound ratio, the denominator always stays the same. Therefore, the value of molecular shows a large fraction of the formula. Given a fuzzy itemset $X$, it is clear that the number of temporal quantitative transactions containing $X$ is always greater than the number of its superset $X^\prime$. Then, the in-equation \textit{tfuubr}$_{X}$ $\ge$ \textit{tfuubr}$_{X^\prime}$ $\ge$ \textit{tfur}$_{X^\prime}$ is always true.
\end{property}

\begin{strategy}
	\label{stra:fuzzyDC}
	\rm Based on Property \ref{pro:fuzzyDC}, if a temporal fuzzy utility upper-bound ratio of a fuzzy itemset is less than the user-defined minimal fuzzy utility threshold $\gamma$, we can directly ignore it and its supersets during the mining process.
\end{strategy}

Since \textit{tfuubr}$_{X}$ is always higher than or equal to \textit{tfuubr}$_{X^\prime}$, we define the symbol $\prec$ as a global descending order of all single items from $I$ to help prune the search space. Thus, a temporal quantitative transaction is ``revised'' when all the contained items in it are \textit{HTFUUIs} and all remaining items are sorted by the global descending order (``$\prec$''). Besides, if a database consists of revised temporal quantitative transactions, the database is considered as \textit{TQD}$^\prime$.

\subsection{TP-table for Fast Locating}

In previous sections, we introduced several basic concepts of temporal fuzzy utility and a revised temporal quantitative database. Since a period may contain zero or one or even more temporal quantitative transactions, it is foreseeable that calculating the temporal fuzzy utility of a fuzzy itemset by scanning over and over again is unacceptable. In this subsection, a transactions-period table (TP-table) structure is proposed to record the relationships between transactions and the periods that include them, which is defined as follows.

\begin{definition}
	\rm Given a temporal quantitative database \textit{TQD}, a TP-table can be created by scanning the \textit{TQD} once, where each temporal quantitative transaction $T_j$ is used as an index to indicate the period $P_i$ that contain it. That is, $P_i$ = \{$T_j$ $|$ $T_j \in P_i$ $\land T_j$ $\in \textit{TQD}$\}.
\end{definition}

Considering Table \ref{tab:database}, a typical example of a TP-table is given in Table \ref{tab:tp}, and a TP-table consists of two fields, including period and \textit{Tids}. In general, a naive method adopts a matrix array (5 $\times$ 10), that the first column stores the period, and the rest columns are used to store \textit{Tids}. Furthermore, bitmaps can also be integrated into the matrix to save memory. What's more, we implement TP-table by hash-map in our work, which accelerates the locating process effectively. According to the utilization of the TP-table, it is clear that it can save memory and runtime consumption because massive unnecessary times of scanning databases are reduced. As databases tend to be large, the improved effect will be more significant.

\begin{table}[h]
	\centering
	\caption{An example of TP-table}
	\label{tab:tp}
		\begin{tabular}{cc}
			\hline
			\textbf{Period} & \textbf{\textit{Tids}} \\
			\hline
			$P_1$ & $T_1$, $T_2$ \\
			$P_2$ & $T_3$, $T_4$ \\
			$P_3$ & $T_5$, $T_6$ \\
			$P_4$ & $T_7$, $T_8$ \\
			$P_5$ & $T_9$, $T_{10}$ \\
			\hline
		\end{tabular}
\end{table}

\subsection{Tighter Remaining Measure}

The mining mechanism of the proposed approach is a depth-first method. The search space is a set-enumeration tree \cite{rymon1992search} with an empty root. The first layer of the tree is filled with single items (1-itemsets) from $I$. Then, 2-itemsets are the child nodes of these 1-itemsets in the second layer. And there are third, fourth layers, and so on in the same manner. Note that the total number of nodes in the search space exceeds $2^{|I|}$. Hence, performing a downward traversal of every branch by a naive method is unrealistic. We next introduce several definitions related to the exploration of fuzzy itemsets in depth-first search space.

\begin{definition}
	\rm In the search space, the set of extension fuzzy items $x_i$ of a fuzzy itemset $X$ is sorted behind $X$ in $I$. In this paper, we adopt the global descending order (``$\prec$'') to obtain large fuzzy itemsets and avoid generating the same fuzzy itemset repeatedly. That is, $E$($X$) = \{$x_i$ $|$ $x_i \in I$ $\land x_c$ $\prec x_i$, $\forall x_c$ $\in X$\}. Then, consider $X$ in a revised temporal quantitative transaction $T_j$, and the set of all fuzzy items after $X$ is named remaining fuzzy items of $X$ and denoted as $T_j$ / $X$ = \{$x_i$ $|$ $x_c$ $\prec x_i$, $\forall x_c$ $\in X$ $\land$ $x_i$ $\in$ $T_j$\}.
\end{definition}

Since the maximal fuzzy utility of an item is the upper-bound in the fuzzy set of the item and a fuzzy itemset does not contain fuzzy items from the same fuzzy set, it is feasible to take the maximal fuzzy utility of an item as the estimation of the extended fuzzy utility of a fuzzy itemset. In other words, the remaining maximal fuzzy utility of a fuzzy itemset represents how much the fuzzy utility of it can increase in the subsequent exploring process.

\begin{definition}
	\rm Given a fuzzy itemset $X$ in a temporal quantitative transaction $T_j$, the utility of remaining fuzzy items $x_i$ $\in$ $T_j$ / $X$ is named remaining maximal temporal fuzzy utility, and defined as \textit{rmtfu}$_{jX}$ = $\sum_{x_i \in T_j / X}$\textit{mfu}$_{ij}$. 
\end{definition}

For example, consider a fuzzy item \textit{B.Low} in the revised temporal quantitative transaction $T_7$. $T_7$ / \textit{B.Low} = \{\textit{B.Middle}, \textit{E.Middle}, \textit{E.High}, \textit{A.Low}, \textit{F.Low}, \textit{C.Low}\}, since \textit{B.Low} and \textit{B.Middle} can not occur in a same fuzzy itemset, thus \textit{rmtfu}$_{T_7, \textit{B.Low}}$ = \textit{mtfu}$_{E, T_7}$ + \textit{mtfu}$_{A, T_7}$ + \textit{mtfu}$_{F, T_7}$ + \textit{mtfu}$_{C, T_7}$ = 5.36 + 12.06 + 2.31 + 1.32 = 21.05.

\subsection{Revised Temporal Fuzzy-List Structure}

In the FUMT algorithm \cite{ye2022fuzzy}, the temporal fuzzy-list structure (\textit{TF-List}) was proposed to store \textit{HTFUUIs} information. The \textit{TF-List} significantly reduces the database scanning times. However, we notice that FUMT takes the maximal fuzzy utility of fuzzy items in the quantitative transaction as the upper-bound, which is too loose to effectively prune the search space. Thus, we modify the \textit{TF-List} based on the remaining measure in our work. We discuss and prove the superiority of the remaining measure in this subsection.

\begin{figure}[!htbp]
	\centering
	\includegraphics[clip,scale=0.4]{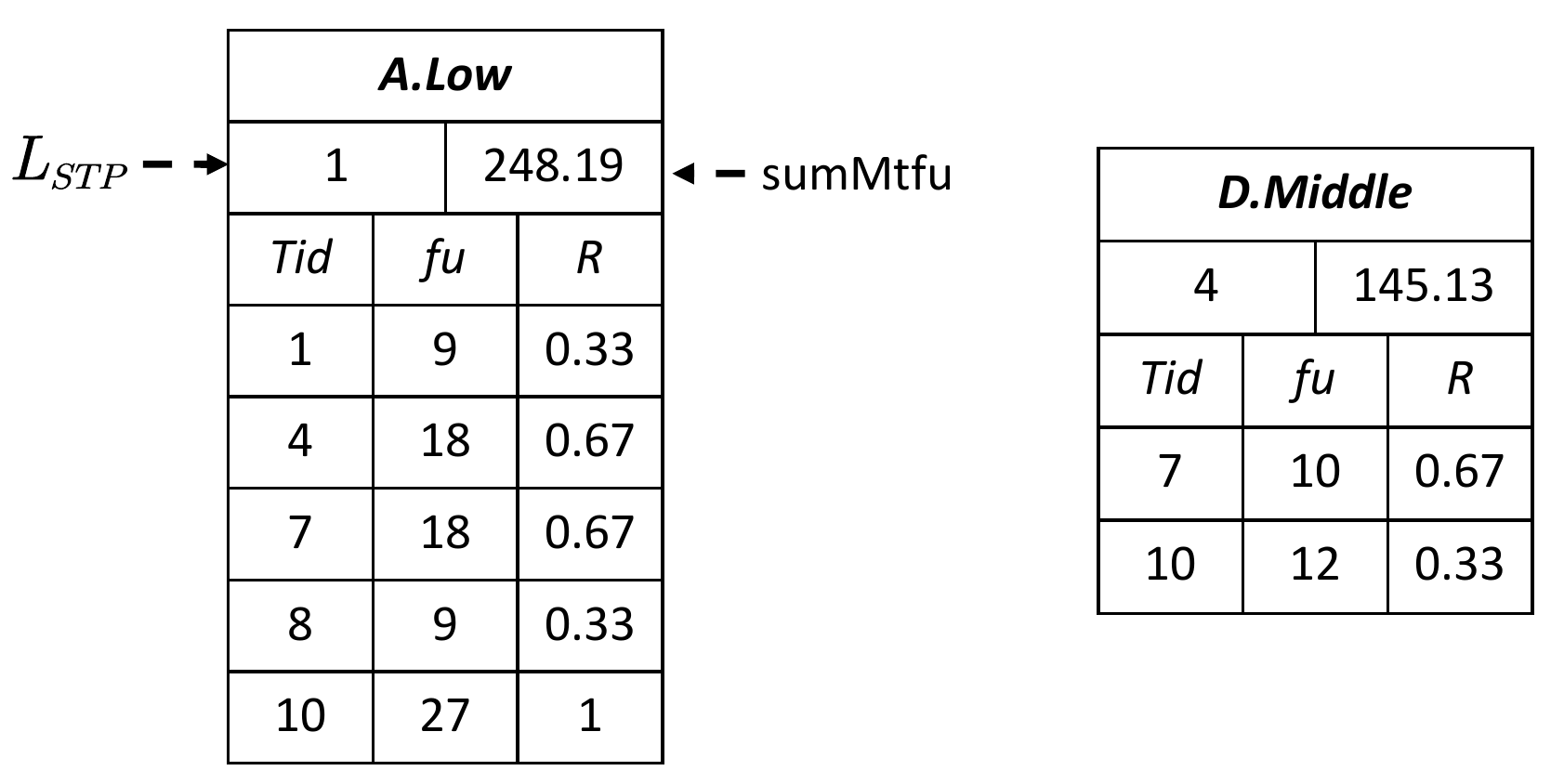}
 	\caption{The temporal fuzzy-lists of some \textit{HTFUUIs}$_1$.}
	\label{fig:tf_list_1}
\end{figure}


\begin{definition}
	\rm A temporal fuzzy-list \cite{ye2022fuzzy} of a fuzzy itemset $X$ is a set of tuples. As shown in Fig. \ref{fig:tf_list_1}, the tuple consists of five elements: 1) transaction identification (\textit{Tid}) shows how many transactions contain $X$; 2) \textit{fu} records the fuzzy utility of $X$ in the corresponding temporal quantitative transaction; 3) $R$ reveals the fuzzy region values of $X$ in different transactions; 4) the latest \textit{STP} ($L_{\textit{STP}}$); and 5) the sum of maximal fuzzy utility of the temporal quantitative transaction (\textit{sumMtfu}).
\end{definition}

After FUMT discovers a complete set of high temporal fuzzy utility upper-bound 1-itemsets (\textit{HTFUUIs}$_1$) by scanning the temporal quantitative database (\textit{TQD}) for the first time, a revised temporal quantitative database (\textit{TQD}$^\prime$) is obtained. FUMT then traverses \textit{TQD}$^\prime$ to construct the temporal fuzzy-lists of \textit{HTFUUIs}$_1$ (\textit{TF-List}$_1$). Take fuzzy item \{\textit{F.Low}\} in Table \ref{tab:database} as an example. Since the \{\textit{F.Low}\} occurs in the transactions $T_1$, $T_2$, $T_5$ and $T_7$, the \textit{STP} of \{\textit{F.Low}\} is $P_1$. From the Table \ref{tab:comparison} we can know the \textit{mtfus} of \{\textit{F.Low}\} are 24.68, 21.71, 52.65, and 41.15, respectively. Then the \textit{sumMtfu} can be calculated as (24.68 + 21.71 + 52.65 + 41.15), which is 140.19. During the process of obtaining the \textit{mtfu} and calculating the \textit{sumMtfu}, it stores the \textit{Tid}, \textit{fu} and \textit{R} in the list. By the same way, the FUMT algorithm constructs the \textit{TF-List}$_1$. In addition, \textit{sumMtfu} is an upper-bound that keeps decreasing. Its value reflects how much the fuzzy utility of fuzzy itemset can increase. Considering the fuzzy 2-itemset \{\textit{B.Low}, \textit{A.Low}\}, the itemset \{\textit{B.Low}, \textit{A.Low}\} appears in $T_7$ and $T_{10}$. The \textit{mtfu} of these two transactions is 41.15 and 103.98, respectively. Therefore, the \textit{sumMtfu} of the itemset \{\textit{B.Low}, \textit{A.Low}\} is 145.13. As for the latest \textit{STP}, since the \textit{STP} of \textit{B.Low} and \textit{A.Low} are both $P_1$ that the latest \textit{STP} of \{\textit{B.Low}, \textit{A.Low}\} is $P_1$. The information for \textit{Tid}, \textit{fu} and \textit{R} is recorded in the same manner.

\begin{figure*}[h]
	\centering 
	\includegraphics[clip,scale=0.4]{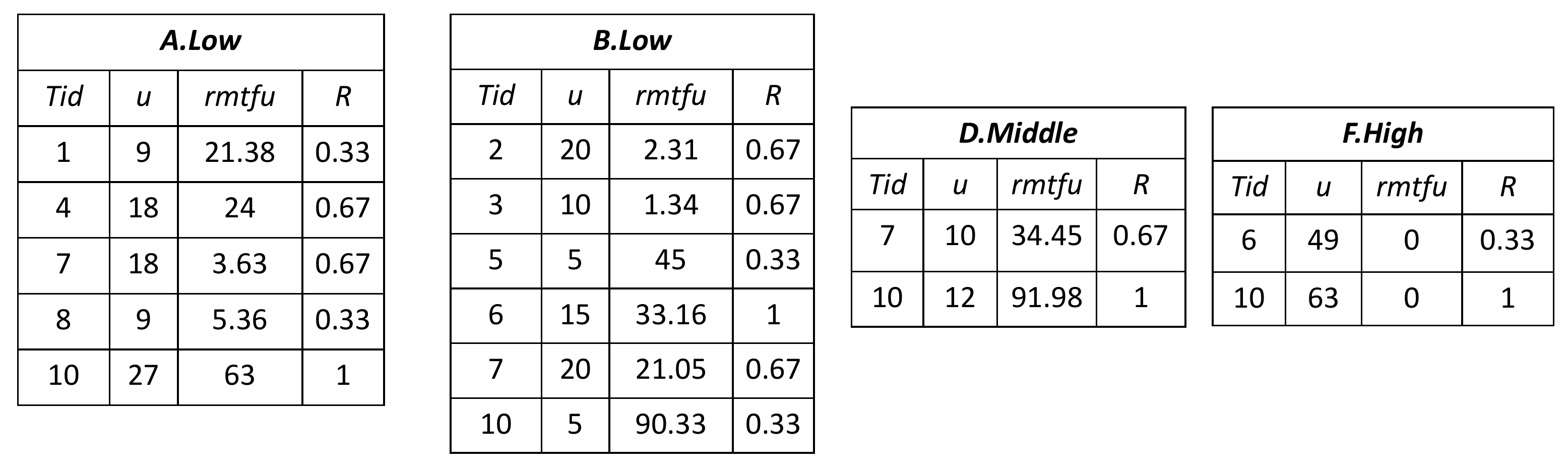}
	\caption{The revised temporal fuzzy-lists of several \textit{HTFUUIs}$_1$.}
	\label{fig:rtf_list_1}
\end{figure*}

However, taking \{\textit{B.Low}, \textit{A.Low}\} as the same sample, the remaining maximal fuzzy utility of it is computed as \{\textit{B.Low}, \textit{A.Low}\}.\textit{rmtfu} = 63, which is far less than its \textit{sumMtfu}. Fig. \ref{fig:rtf_list_1} lists \textit{rmtfu} of other \textit{HTFUUIs}$_1$.

\begin{definition}
	\rm Based on the remaining fuzzy utility definition, we modify the temporal fuzzy-list structure as the revised temporal fuzzy-list (\textit{RTF-List}) structure. An \textit{RTF-List} consists of a set of tuples, and each tuple contains four elements: 1) transaction identification (\textit{Tid}); 2) utility of itemset ($u$); 3) remaining maximal temporal fuzzy utility (\textit{rmtfu}); and 4) fuzzy region value ($R$).
\end{definition}

\begin{figure}[h]
	\centering 
	\includegraphics[clip,scale=0.42]{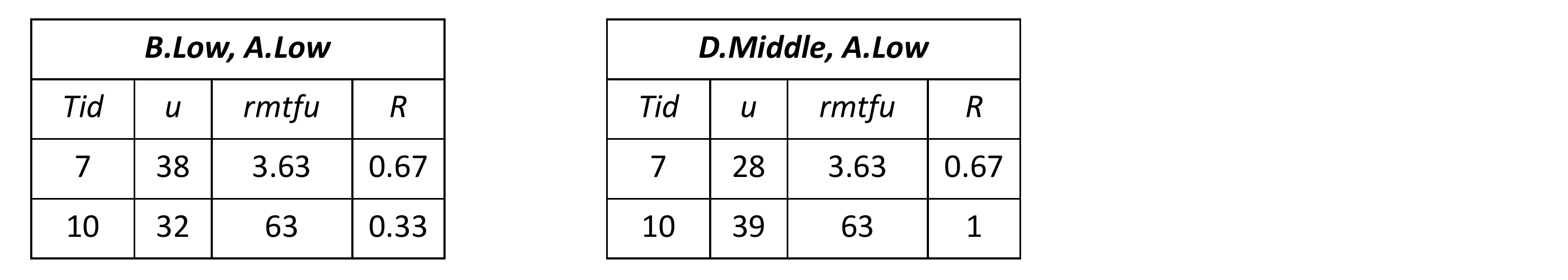}
	\caption{The revised temporal fuzzy-lists of several fuzzy 2-itemsets.}
	\label{fig:rtf_list_2}
\end{figure}

There is no need for a database scan because the proposed method generates fuzzy 2-itemsets by joining RTF-Lists of different low-level fuzzy itemsets. According to Definition \ref{def:fu}, the fuzzy utility of an itemset is utility multiplied by region value. In order to reduce the amount of calculating, the \textit{RTF-List} structure records the utility of an item. We also take \{\textit{B.Low}, \textit{A.Low}\} as illustrations. The common \textit{Tids} in two \textit{RTF-Lists} of \textit{B.Low} and \textit{A.Low} are $T_7$ and $T_{10}$. In other words, the high-level fuzzy itemset \{\textit{B.Low}, \textit{A.Low}\} occurs in these temporal quantitative transactions. Thus, \textit{RTF-List}$_{\{\textit{B.Low}, \textit{A.Low}\}}$ contains two tuples. The \textit{u} of \{\textit{B.Low}, \textit{A.Low}\} in each tuple is the summation of the utility of \textit{B.Low} and \textit{A.Low}. And the \{\textit{B.Low}, \textit{A.Low}\}.\textit{rmtfu} is equal to that of \textit{A.Low} because \textit{A.Low} is a remaining fuzzy item of \textit{B.Low}. Finally, Fig. \ref{fig:rtf_list_2} lists the details of \textit{RTF-List}$_{\{\textit{B.Low}, \textit{A.Low}\}}$ and others.

Then we consider the \textit{RTF-List} of fuzzy $k$-itemset ($k \ge 3$) and take the same manner as described previously. The difference between intersecting \textit{RTF-Lists} of fuzzy ($k$-1)-itemsets and constructing \textit{RTF-List} of fuzzy 2-itemset is that \textit{u} of fuzzy $k$-itemset has double-calculated the fuzzy utility of its prefix fuzzy itemset (\{$x_1$, $\ldots$, $x_{k-2}$\}). Note that this is a miscalculation. For example, in Fig. \ref{fig:rtf_list_2}, the $u$ of the element associated with $T_c$ in \textit{RTF-List}$_{\{\textit{B.Low}, \textit{A.Low}, \textit{F.High}\}}$ is calculated as \{\textit{B.Low}, \textit{A.Low}, \textit{F.High}\}.\textit{u} = \{\textit{B.Low}, \textit{A.Low}\}.\textit{u} + \{\textit{B.Low}, \textit{F.High}\}.\textit{u} - \{\textit{B.Low}\}.\textit{u}.

\begin{definition}
	\rm In \textit{RTF-List} of a fuzzy itemset $X$, the sum of \textit{fu} elements is denoted as $X$.\textit{sumFu} = $\sum_{T_j \in \textit{RTF-List}_X}$\textit{fu}$_{jX}$($X$, $T_j$), and the sum of \textit{rmtfu} elements is denoted as $X$.\textit{sumRmtfu} = $\sum_{T_j \in \textit{RTF-List}_X}$\textit{rmtfu}$_{jX}$($X$, $T_j$).
\end{definition}


\begin{property}
	\label{pro:rmtfu}
	\rm Since the \textit{RTF-List} contains all \textit{rmtfu} information of a fuzzy itemset $X$, \textit{sumRmtfu} can reflect how much the fuzzy utility of $X$ increases in \textit{TQD}. Hence, if $X$\textit{.sumFu} + $X$\textit{.sumRmtfu} $\ge$ $\gamma$ $\times$ $\sum_{T_j \in \textit{LTP}_{\textit{all}}}$\textit{tfu}$_j$ is true, there must exist one or more fuzzy super-itemsets of $X$ are \textit{HTFUIs}.
\end{property}

\begin{strategy}
	\label{stra:rmtfu}
	\rm Given a fuzzy itemset $X$ and its \textit{RTF-List}$_X$. If the condition $X$.\textit{sumFu} $+$ $X$.\textit{sumRmtfu} $<$ $\gamma$ $\times$ $\sum_{T_j \in \textit{LTP}_{\textit{all}}}$\textit{tfu}$_j$ is true, there is no need to explore the search space for $X$.
\end{strategy}

\begin{proof}
    Given a fuzzy itemset $X$ and its extension $X^\prime$ ($X \subset X^\prime$), and for $\forall X^\prime \subseteq T_j$ there is ($X^\prime - X$) = ($X^\prime$ / $X$).

    \begin{tabbing}
        $\because$ $X \subset X^\prime \subseteq T_j$ $\Rightarrow$ ($X^\prime / X$) $\subseteq$ ($T_j / X$),
    \end{tabbing}
    \begin{tabbing}
	    $\therefore$ \textit{fu}$_{jX^\prime}$($X^\prime$, $T_j$) \= 
	    = \textit{fu}$_{jX}$($X$, $T_j$) + \textit{fu}$_{j(X^\prime - X)}$(($X^\prime - X$), $T_j$) \\
	    \> = \textit{fu}$_{jX}$($X$, $T_j$) + \textit{fu}$_{j(X^\prime / X)}$(($X^\prime / X$), $T_j$) \\
	    \> = \textit{fu}$_{jX}$($X$, $T_j$) + $\sum_{x_i \in X^\prime / X}$\textit{fu}$_{jx_i}$($x_i$, $T_j$) \\
	    \> $\le$ \textit{fu}$_{jX}$($X$, $T_j$) + $\sum_{x_i \in T_j / X}$\textit{fu}$_{jx_i}$($x_i$, $T_j$) \\
	    \> $\le$ \textit{fu}$_{jX}$($X$, $T_j$) + $\sum_{x_i \in T_j / X}$\textit{mfu}$_{jx_i}$($x_i$, $T_j$) \\
	    \> = \textit{fu}$_{jX}$ + \textit{rmtfu}$_{jX}$.
    \end{tabbing}
    
    Suppose a \textit{Tid} set of $X$ in list is denoted as $|\textit{RTF-List}_{X}|$, and a \textit{Tid} set of $X^\prime$ in \textit{LTP}$_{X^\prime}$ is defined as $|\textit{LTP}_{X^\prime}|$, then:
    
    \begin{tabbing}
        $\because$ $|\textit{RTF-List}_{X}|$ $\ge$ $|\textit{LTP}_{X^\prime}|$, and \textit{LTP}$_{\textit{all}}$ $\subseteq$ \textit{LTP}$_{X^\prime}$,
    \end{tabbing}
    \begin{tabbing}
	    $\therefore$ \textit{tfur}$_{X^\prime}$ \=
	    = $\sum_{X \in T_j \land T_j \in \textit{LTP}_{X^\prime}}\textit{fu}_{jX^\prime}$ $/$ $\sum_{T_j \in \textit{LTP}_{X^\prime}}\textit{tfu}_{j}$ \\
	    \> $\le$ $\sum_{T_j \in \textit{RTF-List}_{X}}(\textit{fu}_{jX} + \textit{rmtfu}_{jX})$ $/$ $\sum_{T_j \in \textit{LTP}_{X^\prime}}\textit{tfu}_{j}$ \\
	    \> $\le$ $\sum_{T_j \in \textit{RTF-List}_{X}}(\textit{fu}_{jX} + \textit{rmtfu}_{jX})$ $/$ $\sum_{T_j \in \textit{LTP}_{\textit{all}}}\textit{tfu}_{j}$ \\
	    \> = $X.\textit{sumFu} + X.\textit{sumRmtfu}$ $/$ $\sum_{T_j \in \textit{LTP}_{\textit{all}}}\textit{tfu}_{j}$.
    \end{tabbing}
    The proof is done.
\end{proof}

\begin{property}
    \label{pro:rmtfu+}
    \rm Assume that the intersection of two distinct \textit{RTF-Lists} (\textit{RTF-List}$_X$ and \textit{RTF-LIst}$_Y$) is not empty. In fact, Strategy \ref{stra:rmtfu} considers all elements of \textit{RTF-List}$_X$. Thus, we propose a tighter remaining upper-bound than the last one.
\end{property}

\begin{strategy}
    \label{stra:rmtfu+}
    \rm Given two fuzzy itemsets $X$ and $Y$, where $X \not= Y$ and \textit{RTF-List}$_X$ $\cap$ \textit{RTF-List}$_Y$ $\not= \emptyset$. If $X$\textit{.sumFu} + $X$\textit{.sumRmtfu} $-$ $\sum_{Y \not\subseteq T_j \land T_j \in \textit{RTF-List}_{X}}$($X$\textit{.fu} + $X$\textit{.rmtfu}) $<$ $\gamma$ $\times$ $\sum_{T_j \in \textit{LTP}_{\textit{all}}}\textit{tfu}_{j}$, all supersets of \textit{XY} can be pruned safely.
\end{strategy}

\begin{proof}
    Assume the extension of fuzzy itemsets $X$ and $Y$ are $X^\prime$ and $Y^\prime$, respectively. Similarly, the extension of \textit{XY} is \textit{X}$^\prime$\textit{Y}$^\prime$.
    
    \begin{tabbing}
        $\because$ $\sum_{T_j \in \textit{RTF-List}_{X}}$(\textit{fu}$_{jX}$ $+$ \textit{rmtfu}$_{jX}$) \= \\
        = $\sum_{T_j \in \textit{RTF-List}_{X} \land Y \subseteq T_j}$(\textit{fu}$_{jX}$ $+$ \textit{rmtfu}$_{jX}$) \\
        $+$ $\sum_{T_j \in \textit{RTF-List}_{X} \land Y \not\subseteq T_j}$(\textit{fu}$_{jX}$ $+$ \textit{rmtfu}$_{jX}$),
    \end{tabbing}
    \begin{tabbing}
        $\therefore$ $\sum_{T_j \in \textit{RTF-List}_{X} \land Y \subseteq T_j}$(\textit{fu}$_{jX}$ $+$ \textit{rmtfu}$_{jX}$) \= \\
        = $\sum_{T_j \in \textit{RTF-List}_{X}}$(\textit{fu}$_{jX}$ $+$ \textit{rmtfu}$_{jX}$) \\
        $-$ $\sum_{T_j \in \textit{RTF-List}_{X} \land Y \not\subseteq T_j}$(\textit{fu}$_{jX}$ $+$ \textit{rmtfu}$_{jX}$) \\
        = $\sum_{T_j \in \textit{RTF-List}_{\textit{XY}}}$(\textit{fu}$_{j\textit{XY}}$ $+$ \textit{rmtfu}$_{j\textit{XY}}$).
    \end{tabbing}
    
    Assume that $T_p \in \textit{RTF-List}_{\textit{XY}}$, $T_q \in \textit{RTF-List}_{\textit{X}^\prime\textit{Y}^\prime}$, and $T_j \in \textit{RTF-List}_{X}$, thus $T_q \subseteq T_p \subseteq T_j$.
    
    \begin{tabbing}
        $\therefore$ \textit{fu}$_{\textit{X}^\prime\textit{Y}^\prime}$ \=
        = $\sum_{T_q \in \textit{RTF-List}_{\textit{X}^\prime\textit{Y}^\prime}}$\textit{fu}$_{q\textit{X}^\prime\textit{Y}^\prime}$ \\
        \> $\le$ $\sum_{T_p \in \textit{RTF-List}_{\textit{XY}}}$(\textit{fu}$_{p\textit{XY}}$ $+$ \textit{rmtfu}$_{p\textit{XY}}$) \\
        \> =  $\sum_{T_j \in \textit{RTF-List}_{X}}$(\textit{fu}$_{jX}$ $+$ \textit{rmtfu}$_{jX}$) \\
        \> $-$ $\sum_{T_j \in \textit{RTF-List}_{X} \land Y \not\subseteq T_j}$(\textit{fu}$_{jX}$ $+$ \textit{rmtfu}$_{jX}$).
    \end{tabbing}
    
    Based on the proof details of Property \ref{pro:rmtfu}, Property \ref{pro:rmtfu+} is proved.
\end{proof}




\subsection{The TFUM Algorithm}

The pseudocode of TFUM is presented in \textbf{Algorithm} \ref{algo:tfum_algorithm}. It takes three input parameters: 1) a quantitative transaction database that is temporal; 2) a user-specified minimum threshold; and 3) a pre-defined membership function. The TFUM algorithm firstly initializes both the TP-table and high temporal fuzzy utility of fuzzy 1-itemsets as empty (Line 1). In Lines 2--10, TFUM aims to prepare for the subsequent mining process, which includes updating the TP-table and \textit{HTFUUIs}$_1$, adding the start time period of each fuzzy item to \textit{STP-List}, and \textit{STP}$_{\textit{all}}$. In order to make the following discussion easier, we use the symbol $\delta$ to represent $\gamma \times \sum_{T_j \in \textit{LTP}_{\textit{all}}}\textit{tfu}_j$ (Line 11). Since a complete set of high temporal fuzzy utilities of fuzzy 1-itemsets is collected, TFUM will get a global order ``$\prec$'' and a revised temporal quantitative database in Lines 12 and 13. Subsequently, in Line 14, the revised temporal fuzzy-lists of \textit{HTFUUIs}$_1$ will be generated, and then TFUM keeps relevant information about fuzzy itemsets from the revised database in memory. In Line 15, after TFUM has finished initializing the prefix fuzzy itemset $P$ and its list \textit{RTF-List}$_{P}$, TFUM calls a recursive method until no more high temporal fuzzy utility itemsets are generated (Line 16). Finally, a complete set of \textit{HTFUIs} will be output when the TFUM algorithm terminates.

\begin{algorithm}[h]
	\caption{The TFUM algorithm}
	\label{algo:tfum_algorithm}
	\LinesNumbered
	\KwIn{\textit{TQD}: a temporal quantitative transaction database; $\gamma$: a user-specified minimum temporal fuzzy utility threshold; $f(R)$: a pre-defined membership function.}
	\KwOut{\textit{HTFUIs}: a complete set of temporal fuzzy high utility itemsets.}
	
	initialize TP-table, \textit{STP-List} and \textit{HTFUUIs}$_1$ as \textit{null};
	
	\For{\rm temporal transaction $T_j$ $\in$ \textit{TQD}}{
	    update TP-table;
	    
	    \For{\rm  each item $x_i \in T_j$}{
	        
	        calculate \textit{fu} of fuzzy items $x_{il}$ by $f(R)$;
	        
	        \textit{STP-List} $\gets$ the start time period (\textit{STP}$_i$) of $x_{il}$;
	        
	    }
	    
	    compute \textit{tfu} and \textit{mtfu} of $T_j$;
	}
	
	\textit{STP}$_{\textit{all}}$ = \textit{max}$_{TP}$\{\textit{STP-List}\};
	
	set $\delta$ as $\gamma$ $\times$ $\sum_{T_j \in \textit{LTP}_{\textit{all}}}\textit{tfu}_j$;
	
	collect a complete set of \textit{HTFUUIs}$_1$;
	
	get the revised database \textit{TQD}$^\prime$ by global order $\prec$;
	
	scan \textit{TQD}$^\prime$ and construct \textit{RTF-Lists}$_1$ of \textit{HTFUUIs}$_1$;
	
	initialize prefix fuzzy itemset $P$ and \textit{RTF-List}$_P$ as \textit{null};
	
	call \textbf{Miner}($P$, \textit{RTF-List}$_P$, \textit{RTF-List}$_1$, $\gamma$, $\delta$);
	
	\textbf{return} \textit{HTFUIs}
\end{algorithm}

\begin{algorithm}[h]
	\caption{The Miner procedure}
	\label{algo:miner}
	\KwIn{$P$: the prefix fuzzy itemset; \textit{RTF-List}$_P$: a revised temporal fuzzy-list of $P$; \textit{RTF-Lists}: a set of revised temporal fuzzy-lists; $\gamma$: a minimum fuzzy utiltiy threshold; $\delta$: a fuzzy utility upper-bound.}
	
	\For{\rm  each \textit{RTF-List}$_X$ $\in$ \textit{RTF-Lists}}{
		\If{\textit{tfur}$_X$ $\ge$ $\gamma$}{
			$X$ is a \textit{HTFUI};
		}
		
		initialize \textit{exRTF-Lists} as \textit{null};
		
		\If{$X$.\textit{sumFu} + $X$.\textit{sumRmtfu} $\ge$ $\delta$}{
			\For{each \textit{RTF-List}$_Y$ $\in$ \textit{RTF-lists}}{
				
				\textit{RTF-List}$_{\textit{XY}}$ = \textbf{Construct}(\textit{RTF-List}$_P$, \textit{RTF-List}$_X$, \textit{RTF-List}$_Y$);
				
				\If{${\textit{XY}}$.\textit{sumRmtfu} $>$ $\delta$}{
					\textit{exRTF-Lists} += \textit{RTF-List}$_{\textit{XY}}$;
				}
			}
			
			$P$ $\leftarrow$ $P$ $\cup$ $X$;
			
			call \textbf{Miner}($P$, \textit{RTF-List}$_X$, \textit{exRTF-Lists}, $\delta$, $\gamma$);
		}
	}
\end{algorithm}

\begin{algorithm}[h]
	\caption{The Construct procedure}
	\label{algo:construct_lsit}
	
	\KwIn{\textit{RTF-List}$_{P}$: a revised temporal fuzzy-list of prefix fuzzy itemset $P$; \textit{RTF-List}$_{\textit{Px}}$: a revised temporal fuzzy-list of fuzzy itemset \textit{Px}; \textit{RTF-List}$_{\textit{Py}}$: a revised temporal fuzzy-list of fuzzy itemset \textit{Py}.}
	\KwOut{a revised temporal fuzzy-list of high-level fuzzy itemset \textit{Pxy}.}
		
	initialize \textit{RTF-List}$_{\textit{Pxy}}$ as \textit{null};
	
	\textit{Pxy.L}$_{\textit{STP}}$ = \textit{Max}\{\textit{Px.L}$_{\textit{STP}}$, \textit{Py.L}$_{\textit{STP}}$\};
		
	\For{\rm each element \textit{Px}$_e$ $\in$ \textit{RTF-List}$_{\textit{Px}}$}{
		\If{$\exists \textit{Py}_e \in \textit{RTF-List}_{\textit{Py}}$ and $\textit{Px}_e$\textit{.Tid} == $\textit{Py}_e$\textit{.Tid}}{
			
			\textit{Pxy}$_e$\textit{.R} = \textit{Min}\{\textit{Px}$_e$\textit{.R}, \textit{Py}$_e$\textit{.R}\};
			
			\eIf{\textit{RTF-List}$_{P}$ $\not=$ \textit{null}}{
				find $P_e$ $\in$ \textit{RTF-List}$_{P}$ where  $P_e$\textit{.Tid} == $\textit{Px}_e$\textit{.Tid}; \\
				\textit{Pxy}$_e$.\textit{u} = \textit{Px}$_e$.\textit{u} + \textit{Py}$_e$.\textit{u} - \textit{P}$_e$.\textit{u};
			} {
				\textit{Pxy}$_e$.\textit{u} = $\textit{Px}_e$.\textit{u} + $\textit{Py}_e$.\textit{u};
			}
			\textit{Pxy}$_e$ = (\textit{Px}$_e$\textit{.Tid}, \textit{Pxy}$_e$.\textit{u}, \textit{Py}$_e$\textit{.rmtfu}, \textit{Pxy}$_e$\textit{.R});
			
			add \textit{Pxy}$_e$ into \textit{RTF-List}$_\textit{Pxy}$;
		}
	}
	\textbf{return} \textit{RTF-List}$_\textit{Pxy}$
\end{algorithm}

The mining procedure of the TFUM algorithm is shown in \textbf{Algorithm} \ref{algo:miner}. The miner procedure takes a prefix fuzzy itemset $P$, a set of revised temporal fuzzy-lists (including \textit{RTF-List}$_{P}$), the minimum threshold $\gamma$, and a fuzzy utility upper-bound $\delta$\footnote{$\delta$ = $\gamma \times \sum_{T_j \in \textit{LTP}_{\textit{all}}}\textit{tfu}_j$, which is described in the last paragraph.}. The procedure will traversal each \textit{RTF-List}$_{X}$ in the input set \textit{RTF-Lists}. It firstly will check the current fuzzy itemset $X$ is an \textit{HTFUI} or not (Line 2). If yes, $X$ will be added into \textit{HTFUIs} set (Line 3). Then, the procedure checks $X$ whether it can be extended or not in the following steps. The procedure initializes \textit{RTF-List} of the extension fuzzy itemset of $X$ as null, and then assesses the summation of fuzzy utility and remaining maximal temporal fuzzy utility of $X$ is higher than $\delta$ or not (Lines 5 and 6). If the condition is true, the miner procedure will examine \textit{RTF-List} of other fuzzy itemsets $Y$ in \textit{RTF-Lists}, where $X \prec Y$, to determine the super fuzzy itemset \textit{XY} should be generated or not (Lines 7--12). While there is no more remaining fuzzy itemset of $X$ existing, the procedure will add $X$ into $P$ and recursively call \textbf{Algorithm} \ref{algo:miner} itself.

\textbf{Algorithm} \ref{algo:construct_lsit} shows the procedure of \textit{TF-List} construction. The procedure joins \textit{RTF-Lists} of two different fuzzy itemsets and then outputs a novel \textit{RTF-List} of the fuzzy super-itemset. The construct procedure firstly initials \textit{RTF-List}$_{\textit{xy}}$ and \textit{Pxy.L}$_{\textit{STP}}$ in Lines 1 and 2. Then, it joins elements of \textit{RTF-List}$_{\textit{Px}}$ and \textit{RTF-List}$_{\textit{Py}}$ which contain the same \textit{Tids} (Lines 3--15). If the prefix fuzzy itemset is null, then the procedure computes the sum of utility of two elements directly (Line 10); otherwise, the summation value should minus the utility of prefix fuzzy itemset (Line 8). In addition, binary search method can greatly accelerate the intersection process. In Line 12, a new element of \textit{RTF-List}$_{\textit{Pxy}}$ is obtained. And then, the procedure adds it into \textit{RTF-List}$_{\textit{Pxy}}$ (Line 13). When the construct procedure terminates, a novel revised temporal fuzzy utility of fuzzy super-itemset \textit{Pxy} is collected (Line 16).

\section{An Illustrate Example}  \label{sec:example}

To more clearly introduce the TFUM algorithm, in this section, a simple example is given to illustrate how to find temporal fuzzy itemsets from a temporal quantitative database (\textit{TQD}), as shown in Table \ref{tab:database}. There are ten transactions, and these ten transactions are attributed to five periods. The six items in \textit{TQD} are denoted $A$ to $F$, respectively. In addition, the external utility of all items is listed in Table \ref{tab:profit}. The membership function is shown in Fig. \ref{fig:defMembership}. In this example, the minimum threshold $\gamma$ = 20\%. The detailed process of the TFUM algorithm is as follows.


\begin{enumerate}[Step 1:]
    \item To discover the temporal fuzzy utility itemsets, TP-table (Table \ref{tab:tp}) stores the period information of each transaction.
    
    \item While traversing each transaction in \textit{TQD}, the algorithm calculates the fuzzy utility of each item with the given membership function. Then \textit{mtfu} is the sum of all maximum fuzzy utilities in this transaction, and \textit{tfu} is the sum of all fuzzy utilities. For instance, according to the membership function, the transaction $T_7$ can be transformed to \{\textit{A.Low}, \textit{B.Low}, \textit{B.Middle}, \textit{C.Low}, \textit{D.Low}, \textit{D.Middle}, \textit{E.Middle}, \textit{E.High}, \textit{F.Low}\}. Since the fuzzy utility of \textit{B.Low} is larger than that of \textit{B.Middle}, the algorithm takes \textit{B.Low} but not \textit{B.Middle} into account when calculating \textit{mtfu}. The same manner can be done for all transactions in \textit{TQD} and the results are shown in Table \ref{tab:comparison}.
    
    \item In addition, in order to find \textit{STP}$_{all}$ of all fuzzy items, a map \textit{STP-List} records \textit{STP} of each fuzzy item. From the \textit{STP-List}, TFUM finds out the latest \textit{STP} (i.e., \textit{STP}$_{all}$). Since \textit{STP}$_{all}$ is $T_{10}$ which belongs to period $P_5$, \textit{LTP}$_{all}$ is the last period of \textit{TQD} (i.e., $P_5$). 
    
    \item For computing the global order, TFUM utilizes a list structure to record all transactions where the item is located, then according to the \textit{tfuubr} to confirm the order. For instance, item $A$ appears in transactions \{$T_1$, $T_4$, $T_7$, $T_8$, $T_9$, $T_{10}$\} and item $F$ appears in \{$T_1$, $T_2$, $T_5$, $T_6$, $T_7$, $T_{10}$\}. Then the \textit{tfuubr} of $A$ and $F$ is 1.71 and 1.69, respectively. Therefore, $A$ $\prec$ $F$. The algorithm sorts the items in transactions using the global order. After revising all transaction in \textit{TQD}, we will get a \textit{TQD}$^\prime$, as shown in Table \ref{tab:revDatabase}.
    
    \item TFUM scans \textit{TQD}$^\prime$ to construct the \textit{RTF-List}$_1$ for each fuzzy item. Take \textit{A.Low} in Table \ref{tab:revDatabase} as an example. The item $A$ appears in six transactions, but \textit{A.Low} appears in only five transactions $T_1, T_4, T_7, T_8$, and $T_{10}$. According to the positions where \textit{A.Low} is located, the \textit{rmtfu} of \textit{A.Low} can be calculated. Meanwhile, store the utility ($u$) and region value ($R$) of \textit{A.Low} in each transaction, respectively. The \textit{RTF-List}$_{A.Low}$ is shown in Fig \ref{fig:rtf_list_1}.
\end{enumerate}

From now on, since the \textit{RTF-Lists} of all fuzzy 1-itemsets have been constructed, thus invoking the \textit{Miner} procedure to mine all temporal fuzzy utility itemsets. The items in \textit{TQD}$^\prime$ have been sorted, and the mining order is consistent with this order. This will ensure the correctness and completeness of the mining results. Take the temporal fuzzy utility itemsets with the prefix \textit{D.Middle} as an example. The details of the \textit{Miner} procedure are described as below.

\begin{enumerate}[MStep 1:]
    \item In the \textit{Miner} procedure, the algorithm calculates the \textit{tfur}$_{\textit{D.Middle}}$ at first. For \textit{D.Middle}, the \textit{tfur} can be calculated as ($\textit{fu}_{T_7}$ + $\textit{fu}_{T_{10}}$)/($\textit{tfu}_{T_7}$ + $\textit{tfu}_{T_8}$+ $\textit{tfu}_{T_9}$ + $\textit{tfu}_{T_{10}}$) = (6.7 + 3.96)/(53.69 + 38.66 + 69 + 103.98), which is 4.02\%. Since 4.02\% is less than 20\%, \textit{D.Middle} is not a \textit{HTFUI}.
    
    \item Next, it is easy to compute the \textit{tfuubr} of \textit{D.Middle} since the \textit{RTF-List}$_{\textit{D.Middle}}$ had already stored the needed information. The \textit{tfuubr} of \textit{D.Middle} can be calculated as ((6.7 + 34.45) + (12 + 91.98)) / (69 + 103.98) = 83.9\%, which is larger than $\gamma$. This means \textit{D.Middle} can be extended.
    
    \item Table \ref{tab:2_P_D.Middle} shows the results of constructing the \textit{RTF-List}$_2$ with \textit{D.Middle} as prefix. From the value of \textit{tfuubr}, we know that \{\textit{D.Middle, B.Low}\}, \{\textit{D.Middle, B.Middle}\}, \{\textit{D.Middle, E.Low}\}, \{\textit{D.Middle, A.Low}\} and \{\textit{D.Middle, F.High}\} can be extended. Then, TFUM continues to call the \textit{Miner} procedure for discovering high temporal fuzzy utility itemsets.
    
    \item Finally, the whole procedure is completed and the seven high temporal fuzzy utility itemsets in Table \ref{tab:HTFUIs} are output to users.
\end{enumerate}


\begin{table}[!ht]
    \centering
	\caption{A revised temporal quantitative database}
	\label{tab:revDatabase}
		\begin{tabular}{ccc}
			\hline
			\textbf{Period} & \textbf{\textit{Tid}} & \textbf{Revised Transaction (item, quantity)} \\
			\hline
			$P_1$ & $T_1$ & ($E$, 1) ($A$, 1) ($F$, 2) ($C$, 3)\\
			$P_1$ & $T_2$ & ($D$, 3) ($B$, 4) ($F$, 1)\\
			$P_2$ & $T_3$ & ($D$, 1) ($B$, 2) ($E$, 2)\\
			$P_2$ & $T_4$ & ($D$, 1) ($E$, 3) ($A$, 2) ($C$, 6)\\
			$P_3$ & $T_5$ & ($D$, 3) ($B$, 1) ($F$, 3) ($C$, 6)\\
			$P_3$ & $T_6$ & ($B$, 3) ($E$, 1) ($F$, 7)\\
			$P_4$ & $T_7$ & ($D$, 5) ($B$, 4) ($E$, 8) ($A$, 2) ($F$, 1) ($C$, 1)\\
			$P_4$ & $T_8$ & ($B$, 6) ($E$, 1) ($A$, 1) ($C$, 2)\\
			$P_5$ & $T_9$ & ($D$, 3) ($A$, 7)\\
			$P_5$ & $T_{10}$ & ($D$, 6) ($B$, 1) ($E$, 1) ($A$, 3) ($F$, 9)\\
			\hline
		\end{tabular}
\end{table}

\begin{table}[h]
	\centering
	\caption{The results of 2-itemsets with \textit{D.Middle} as prefix}
	\label{tab:2_P_D.Middle}
		\begin{tabular}{lll}
			\hline
			\textbf{\textit{Pattern}} & \textbf{\textit{tfur}} & \textbf{\textit{tfuubr}}\\
			\hline
			$\{$\textit{D.Middle}, \textit{B.Low}$\}$ & 10.03\% & 79.25\% \\
			$\{$\textit{D.Middle}, \textit{B.Middle}$\}$ & 6.05\% & 49.5\% \\
			$\{$\textit{D.Middle}, \textit{E.Low}$\}$ & 2.48\% & 54.51\% \\
			$\{$\textit{D.Middle}, \textit{E.Middle}$\}$ & 2.24\% & 12.50\% \\
			$\{$\textit{D.Middle}, \textit{E.High}$\}$ & 4.55\% & 12.5\% \\
			$\{$\textit{D.Middle}, \textit{A.Low}$\}$ & 21.77\% & 71.91\% \\
			$\{$\textit{D.Middle}, \textit{F.Low}$\}$ & 2.11\% & 4.01\% \\
			$\{$\textit{D.Middle}, \textit{F.High}$\}$ & 43.36\% & 43.36\% \\
			$\{$\textit{D.Middle}, \textit{C.Low}$\}$ & 1.74\% & 2.67\% \\
			\hline
		\end{tabular}
\end{table}


\section{Experimental Results}
\label{sec:experimental}

The experimental algorithms (ATTFUM \cite{hong2022one}, FUMT \cite{ye2022fuzzy}, and TFUM) are all implemented in the Java language, and we conducted the experiments on a computer with an Intel Core 3.0 GHz processor with 16 GB of RAM and a Windows 10 64-bit operating system.

\textbf{Dataset description}. Four different datasets (BMSPOS, Retail, Mushroom, and T40I10D100k) were used to evaluate the performance of the tested algorithms. All the datasets can be easily downloaded from the open source library\footnote{SPMF: \url{http://www.philippe-fournier-viger.com/spmf/index.php}}. The characteristics of databases are shown in Table \ref{tab:experiments}. However, two common features of four datasets are that 1) the quantity range of each item is [1, 6]; and 2) they cannot be used for mining high temporal fuzzy itemsets directly since they only offer utilities without periods. Therefore, in order to test the influence of the number of time periods, each dataset was randomly assigned to 1, 2, or 4 time periods, respectively. In addition, all experiments were executed three times independently. The ``Runtime'' and ``Memory'' respectively indicate the average running time and memory consumption of each experiment under the various thresholds ($\gamma$). Finally, we suppose the algorithm is terminated if its running time exceeds over 3,600 seconds, and we use the symbol ``-'' to represent this in tables.

\begin{table}[!ht]
	\fontsize{8pt}{11pt}\selectfont
	\caption{Information of experimental datasets}
	\label{tab:experiments}
	\setlength{\tabcolsep}{3.8mm}{
		\begin{tabular}{crrrc}
			\hline
			\textbf{Dataset} & \#\textbf{Trans} & \#\textbf{Items} & \#\textbf{AvgLen} & \textbf{Type} \\
			\hline
			BMSPOS & 515,366 & 1,656 & 6.51 & sparse \\ 
			Retail & 88,162 & 16,470 & 10.3 & sparse \\
			Mushroom & 8,124 & 119 & 23.0 & dense \\
			T40I10D100k & 10,000 & 942 & 39.6 & dense \\
			\hline
		\end{tabular}
	}
\end{table}

\begin{table}[!ht]
	\fontsize{8pt}{11pt}\selectfont
	\caption{The peak memory of tested algorithms}
	\label{tab:peak_memory}
	\setlength{\tabcolsep}{5.2mm}{
		\begin{tabular}{crrr}
			\hline
			\textbf{Dataset} & \textbf{ATTFUM} & \textbf{FUMT} & \textbf{TFUM} \\
			\hline
			BMSPOS$_{\textit{P4}}$ & 1196 & 1015 & \textbf{945} \\ 
			Retail$_{\textit{P4}}$ & 679 & 579 & \textbf{183} \\
			Mushroom$_{\textit{P4}}$ & 203 & 198 & \textbf{174} \\
			T40I10D100k$_{\textit{P4}}$ & 2348 & \textbf{1004} & 1214 \\
			\hline
		\end{tabular}
	}
\end{table}

\begin{figure}[!hbt]
	\centering
	\includegraphics[scale=0.42]{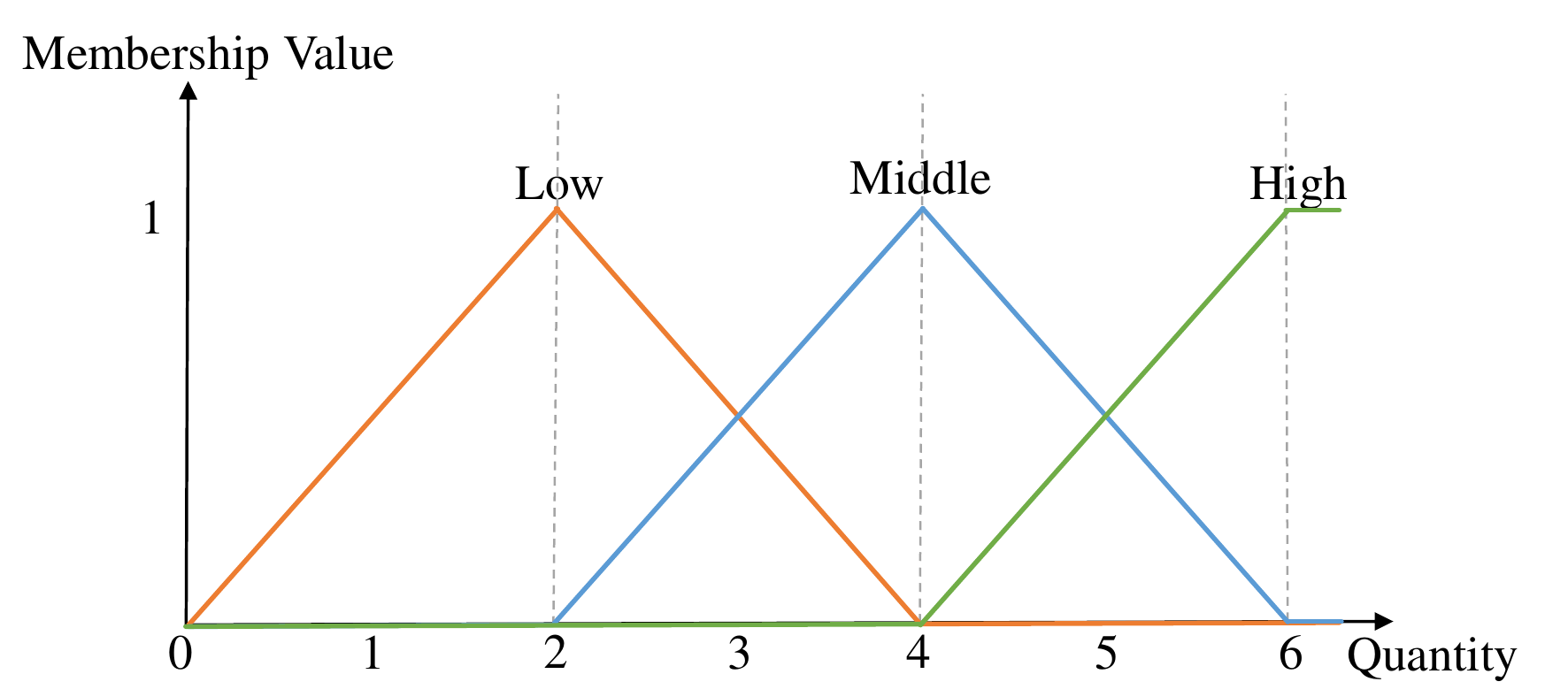}
	\caption{The experimental membership function.}
	\label{fig:expMembership}
\end{figure}

\textbf{Membership function}. For convenience, as shown in Fig. \ref{fig:expMembership}, we supposed all items have the same membership function in our experiments. There are three fuzzy regions (\textit{Low}, \textit{Middle}, and \textit{High}) taken into account.

\begin{figure*}[ht]
	\centering
	\includegraphics[scale=0.6, trim=30 0 0 0]{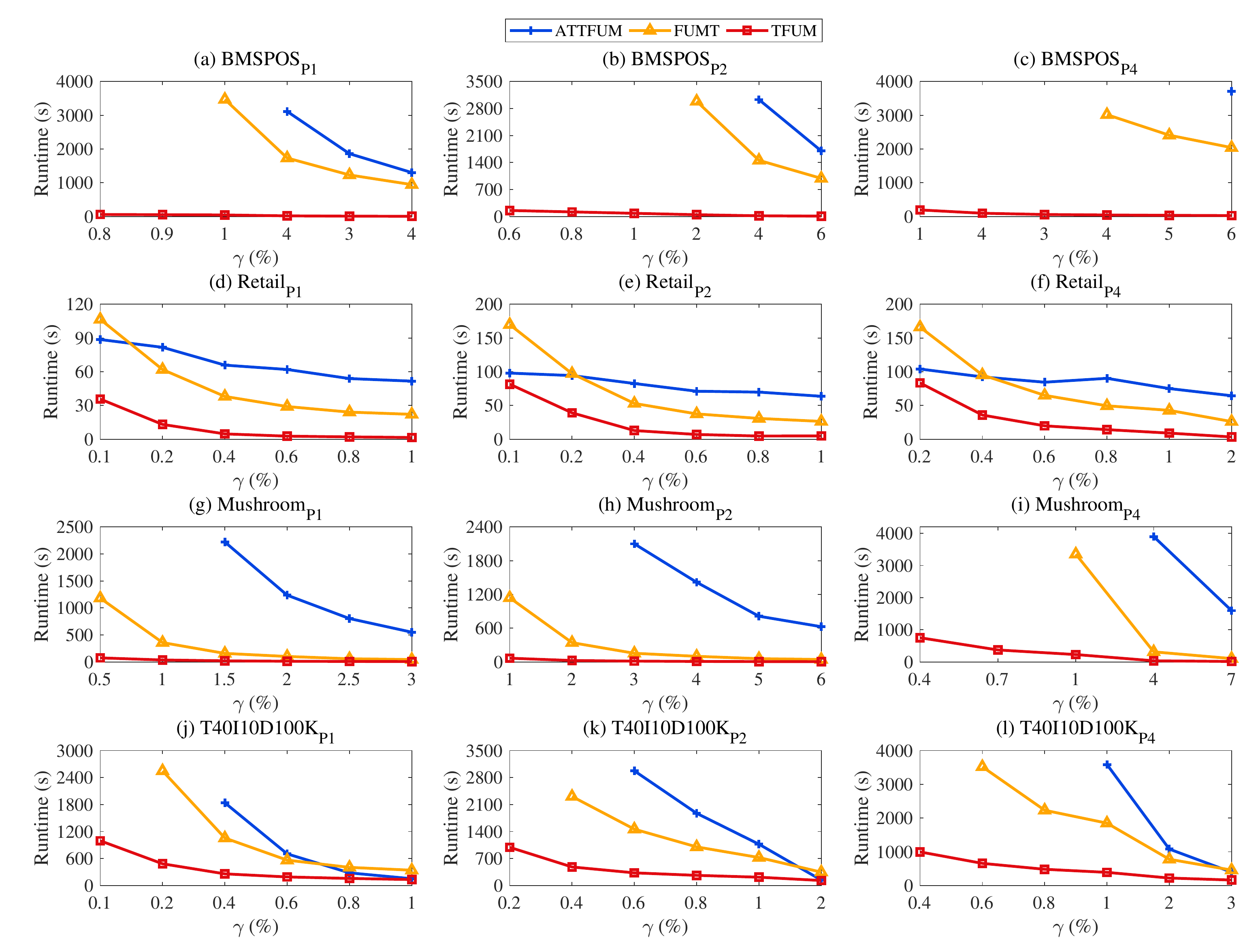}
	\caption{The runtime usage on different periods and thresholds.}
	\label{fig:compare_runtime}
\end{figure*}

\begin{table*}[!htbp]
	\fontsize{8pt}{12pt}\selectfont
	\centering
	\caption{The number of candidates generated by three algorithms}
	\label{tab:candidates}
	\setlength{\tabcolsep}{6mm}{
	\begin{tabular}{ccrrrrrr}
		\hline\hline
		\multirow{2}{*}{\textbf{Dataset}} & \multirow{2}{*}{\textbf{Notation}} & \multicolumn{6}{c}{\textbf{\# Candidate under various thresholds}} \\ \cline{3-8}
		
		& & \textit{test}$_1$ & \textit{test}$_2$ & \textit{test}$_3$ & \textit{test}$_4$ & \textit{test}$_5$ & \textit{test}$_6$ \\ 
		\hline
													 & $\gamma$ & 0.6\% & 0.8\% & 1\% & 2\% & 4\% & 6\% \\
													 & \textbf{ATTFUM} & - & - & - & - & 627 & 242 \\
		\textbf{BMSPOS}$_{\textit{P2}}$ & \textbf{FUMT} & - & - & - & 2,510 & 585 & 238 \\
													 & \textbf{TFUM} & 19,769 & 10,765 & 6,945 & 1,811 & 522 & 267 \\
		\hline
		
													   & $\gamma$ & 1\% & 2\% & 3\% & 4\% & 5\% & 6\% \\
													   & \textbf{ATTFUM} & - & - & 285,203 & 137,647 & 64,054 & 40,047 \\
		\textbf{Mushroom}$_{\textit{P2}}$ & \textbf{FUMT} & 4,013,164 & 616,261 & 182,093 & 90,164 & 41,560 & 26,468 \\
													   & \textbf{TFUM} & 1,181,227 & 209,004 & 75,233 & 35,015 & 19,769 & 12,229 \\
		\hline\hline
	\end{tabular}
	}
\end{table*}

\begin{figure*}[ht]
	\centering
	\includegraphics[scale=0.6, trim=30 0 0 0]{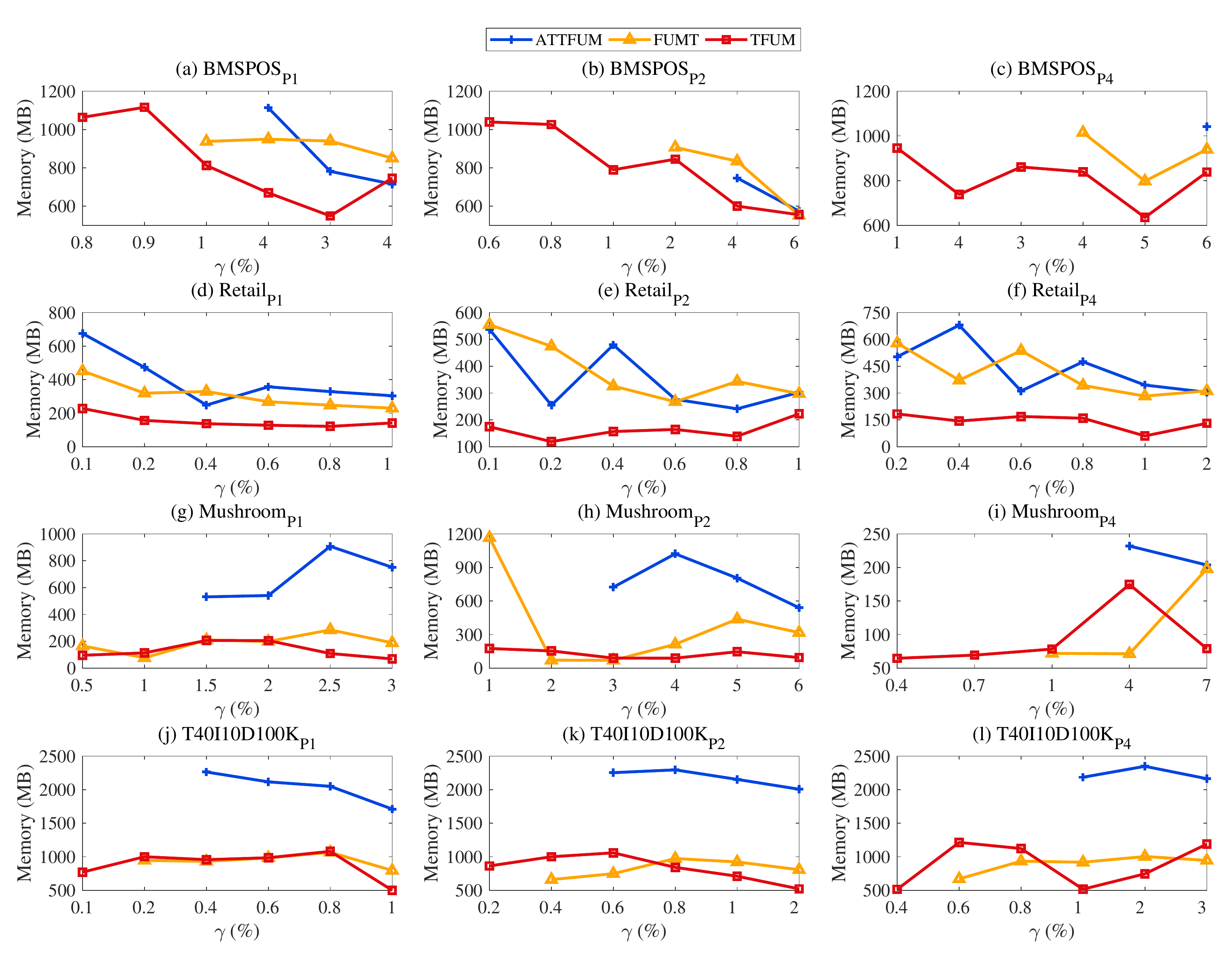}
	\caption{The memory consumption on different periods and thresholds.}
	\label{fig:compare_memory}
\end{figure*}

\begin{table*}[!htbp]
	\fontsize{8pt}{12pt}\selectfont
	\centering
	\caption{The prune ratio of FUMT and TFUM algorithms}
	\label{tab:prune_ratio}
	\setlength{\tabcolsep}{6.3mm}{
		\begin{tabular}{ccrrrrrr}
		\hline\hline
		\multirow{2}{*}{\textbf{Dataset}} & \multirow{2}{*}{\textbf{Notation}} & \multicolumn{6}{c}{\textbf{\# Prune ratio under various thresholds}} \\ \cline{3-8}
		
		& & \textit{test}$_1$ & \textit{test}$_2$ & \textit{test}$_3$ & \textit{test}$_4$ & \textit{test}$_5$ & \textit{test}$_6$ \\ 
		\hline
									   				 & $\gamma$ & 0.6\% & 0.8\% & 1\% & 2\% & 4\% & 6\% \\
		\textbf{BMSPOS}$_{\textit{P2}}$ & \textbf{FUMT} & - & - & - & 95.79\% & 96.38\% & 97.37\% \\
									   				 & \textbf{TFUM} & 98.03\% & 97.98\% & 97.9\% & 97.82\% & 97.1\% & 96.57\% \\
		\hline
		
													   & $\gamma$ & 1\% & 2\% & 3\% & 4\% & 5\% & 6\% \\
		\textbf{Mushroom}$_{\textit{P2}}$ & \textbf{FUMT} & 72.19\% & 76.85\% & 78.16\% & 79.42\% & 80.35\% & 81.47\% \\
													   & \textbf{TFUM} & 91.82\% & 92.75\% & 93.09\% & 93.3\% & 93.38\% & 93.51\% \\
        \hline\hline
		\end{tabular}
	}
\end{table*}


\subsection{Runtime Analysis}

From Fig. \ref{fig:compare_runtime} we can clearly see that the mining performance of ATTFUM and FUMT is degraded consequently while $\gamma$ decreases or the number of time periods increases. For example, in Figs. \ref{fig:compare_runtime} from (g) to (i), FUMT consumes nearly 350 seconds, 1100 seconds, and 3300 seconds in different periods, respectively, when $\gamma$ = 1\%. The runtime cost of ATTFUM increases from 393 seconds to 3584 seconds, while $\gamma$ decreases from 3\% to 1\% in Fig. \ref{fig:compare_runtime}(l). Besides, ATTFUM is always timeout in most cases (especially in Fig. \ref{fig:compare_runtime}(c)). The reason we suppose is that ATTFUM consumes massive runtime in traversing subtrees. In addition, it can be observed that the proposed TFUM algorithm outperforms the other experimental algorithms. For instance, on the BMSPOS dataset, TFUM is up to two orders of magnitude faster than FUMT and ATTFUM. This is reasonable since the TFUM algorithm adopts the TP-table and lists to locate key information in a short time. In the same type of datasets (dense or sparse), the remaining measure is capable of further pruning and then fast discovering \textit{HTFUIs}, when the average length of transactions are shorter. As shown in Fig. \ref{fig:compare_runtime}(h) (Mushroom dataset, two periods, and $\gamma$ = 1\%), TFUM takes only 66 seconds, but FUMT requires 1140 seconds. In Fig. \ref{fig:compare_runtime}, there are many sub-graphs that reveal the much larger gap between the novel algorithm and other algorithms.

\subsection{Memory Usage Analysis}

As illustrated in Fig. \ref{fig:compare_memory}, in general, the memory consumption of the proposed algorithm (i.e., the red line) is usually below that of other experimental algorithms. There are a lot of cases where lines have intersected in Fig. \ref{fig:compare_memory}(a) to (l). For example, on Retail dataset with four time periods, the memory cost of FUMT is higher than that of ATTFUM when $\gamma$ = 0.6\%, but the opposite is true if $\gamma$ is 0.8\%. However, it can be observed that the runtime consumption gap between two algorithms is becoming larger. We suppose this is a trade-off case. In particular, FUMT costs less memory than that of ATTFUM on dense datasets (Figs. \ref{fig:compare_memory}(g)--(l)). In the meantime, the most interesting aspect of Fig. \ref{fig:compare_memory} is that the TFUM algorithm is stabler than others. The reason is that the adopted prune strategies by three algorithms cause a great difference in the number of visited nodes. Table \ref{tab:peak_memory} lists the peak memory of three algorithms on four datasets, and we use bold to highlight the minimal values. As we described in previous content, the more periods, the higher the memory consumption. In fact, the peak memory of ATTFUM and FUMT on the same datasets is comparable, except for T40I10D100k$_{\textit{P4}}$. It is clear that the list-based algorithms perform better than the tree-based algorithms in terms of memory usage. The reason is that the lists compress major information instead of maintaining the dataset in memory.

\begin{figure*}[h]
	\centering
	\includegraphics[height=8cm,width=18cm,scale=0.4]{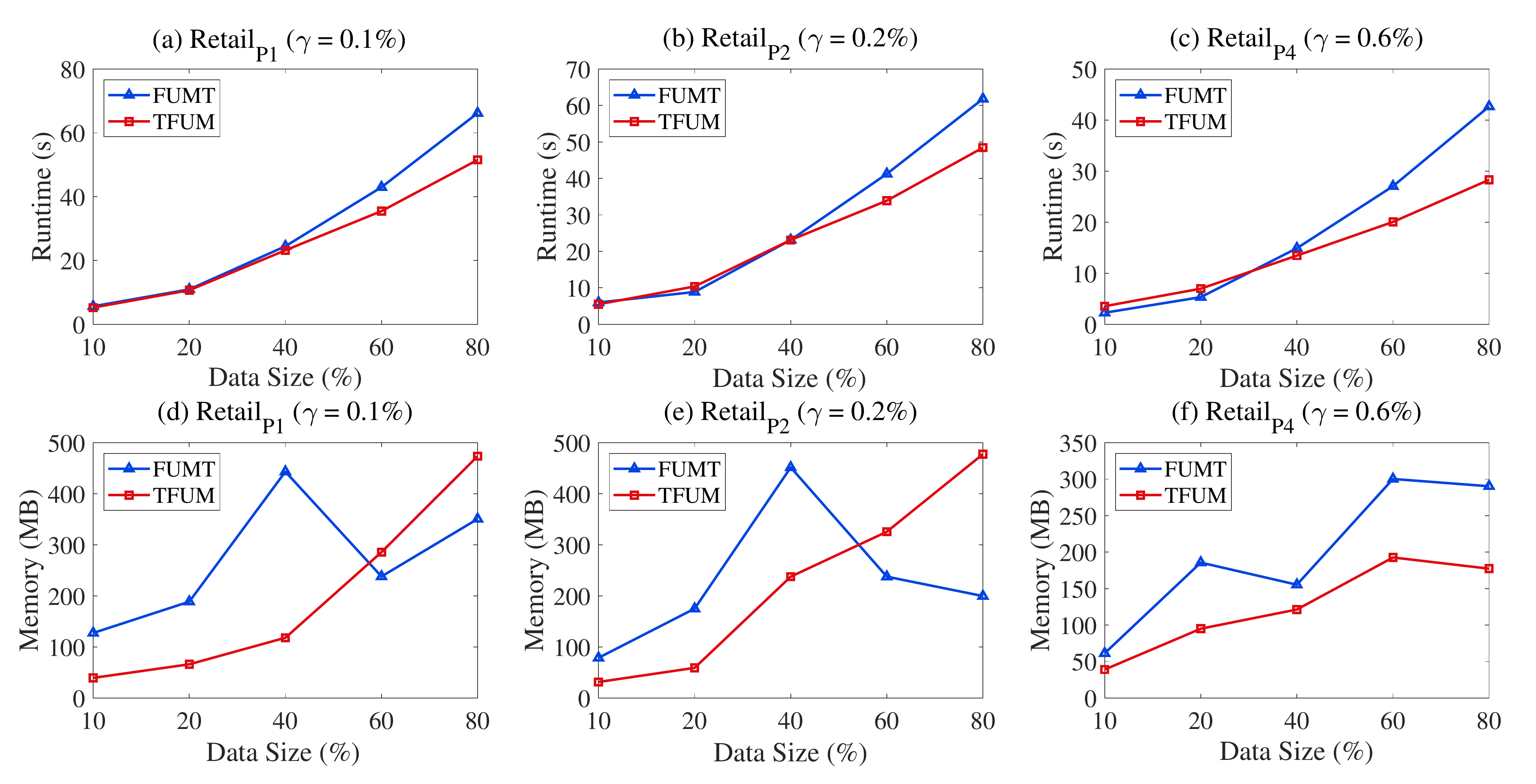}
	\caption{The scalability evaluation on different dataset sizes.}
	\label{fig:scalability_size}
\end{figure*}

\begin{figure}[h]
	\centering
	\includegraphics[trim=40 0 0 0, scale=0.2]{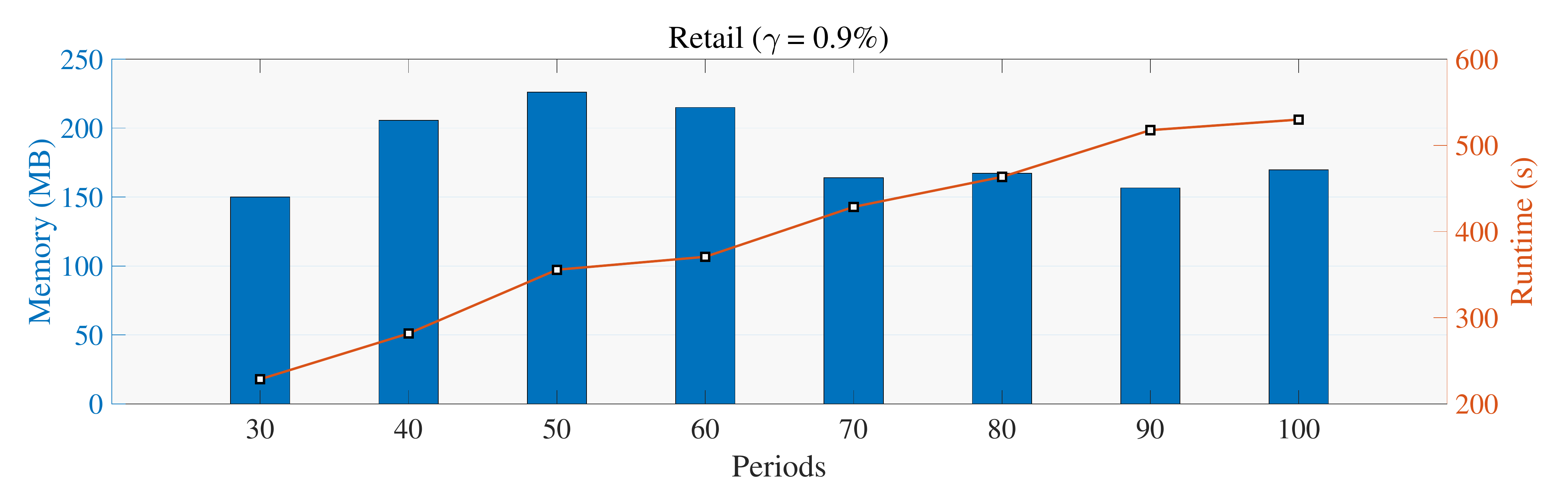}
	\caption{The scalability evaluation on different periods.}
	\label{fig:scalability_period}
\end{figure}

\subsection{Candidate Comparison Analysis}

In this subsection, we compare the amount of candidates generation of experimented algorithms on BMSPOS$_{\textit{P2}}$ and Mushroom$_{\textit{P2}}$ datasets (Table \ref{tab:candidates}). Due to the execution time of ATTFUM exceeds 4900 seconds when $\gamma$ is 2\% on BMSPOS$_{\textit{P2}}$, we can infer that ATTFUM will be over time following $\gamma$ decreases. Considering the Mushroom$_{\textit{P2}}$ dataset, the total number of candidates generation of FUMT is about three times more than that of TFUM. We suppose Properties \ref{pro:rmtfu} and \ref{pro:rmtfu+} play an important role during searching high level fuzzy itemsets. Compared to the naive maximal value upper-bound, the remaining upper bounds are tighter and easily computed (i.e., intersection). Moreover, we also evaluate the prune effect of two list-based algorithms. The formulation is defined as \textit{prune ratio} = $\frac{|\textit{visited nodes} - \textit{candidates}|}{|\textit{candidates}|}$. A fuzzy itemset is assumed as a candidate if it meets Properties \ref{pro:fuzzyDC}, \ref{pro:rmtfu} and \ref{pro:rmtfu+}. Table \ref{tab:prune_ratio} illustrates that the average \textit{prune ratio} of TFUM on BMSPOS$_{\textit{P2}}$ and Mushroom$_{\textit{P2}}$ are 97.6\% and 92.9\%, respectively. In conclusion, the novel algorithm has a higher prune ratio than the benchmark, especially on dense datasets.

\subsection{Scalability Analysis}

We also conducted several experiments to evaluate the scalability of our proposed algorithm on the Retail dataset. In Fig. \ref{fig:scalability_size}, $\gamma$ is assumed to be fixed and the time periods are 1, 2, and 4, respectively. Due to the tested results of 100\% of datasets have already been illustrated in Figs. \ref{fig:compare_runtime} and \ref{fig:compare_memory}, we only conduct 10\%, 20\%, 40\%, 60\%, and 80\% of Retail as experimental datasets. From Figs. \ref{fig:scalability_size}(a) to (f), the runtime and memory consumption steadily increase as the tested data size increases. In addition, TFUM also performs better than FUMT under the same conditions. Then, we evaluate the performance of the TFUM algorithm in different time periods. In Fig. \ref{fig:scalability_period}, we continue to use Retail as experimental dataset and $\gamma$ is set to 0.9\%. Besides, each transaction in Retail will be randomly assigned a period within the given range. And it is possible that a period does not contain any transactions. As can be seen from Fig. \ref{fig:scalability_period}, with the increment of the number of periods, the peak memory usage of TFUM never exceeds 250 MB, and the runtime consumption increases linearly. All in all, we can conclude that the novel algorithm has good scalability performance.

\section{Conclusions and Future Works}
\label{sec:conclusion}

This paper provided a new fuzzy utility mining method called TFUM for efficiently mining HTFUIs from temporal quantitative databases. TFUM used a novel and compact fuzzy-list and TP-table for storing temporal fuzzy information. The novel fuzzy-lists utilize the remaining measure, which makes for tighter upper-bounds than before. Our experimental evaluation of the novel algorithm reveals promising results over the state-of-the-art algorithms on sparse and dense datasets. Extensive experiments show that the adopted remaining measure can efficiently limit the search space of mining and has a high prune ratio on dense datasets. The runtime performance improvements were observed to be much lower for almost all the baseline algorithms. As the threshold was raised, the runtime consumption decreased steadily. Meanwhile, the peak memory requirements in all experiments were also less than those of other tested algorithms. In addition, we also conducted some experiments in terms of scalability. The final results show the proposed algorithm has good scalability no matter the different data sizes or periods. As a part of future work, we intend to consider further reducing the number of visited nodes during the mining process. Other interesting ideas can also be applied to time series forecasting, spatio-temporal applications, and other interesting work.



\bibliographystyle{IEEEtran}
\bibliography{tfum}

\end{document}